\documentclass[graybox, nosecnum]{svmult}

\usepackage{mathptmx}       
\usepackage{helvet}         
\usepackage{courier}        
\usepackage{type1cm}        
\usepackage{makeidx}         
\usepackage{graphicx}        
\usepackage{multicol}        
\usepackage[bottom]{footmisc}
\usepackage{hyperref}        
\usepackage{soul}            

\usepackage{bm}
\usepackage{amsmath}
\usepackage{amssymb}

\makeindex             

\begin{document}
\title*{Theory of kaon-nuclear systems}
\author{Tetsuo Hyodo \thanks{corresponding author} and Wolfram Weise}
\institute{Tetsuo Hyodo \at Tokyo Metropolitan University, Hachioji 192-0397,  Tokyo,  Japan\\\email{hyodo@tmu.ac.jp}
\and Wolfram Weise \at Physics Department,Technical University of Munich, D-85748 Garching, Germany\\ \email{weise@tum.de}}
%
%
\maketitle
\abstract{
The strong interaction between an antikaon and a nucleon is at the origin of various interesting phenomena in kaon-nuclear systems. In particular, the interaction in the isospin $I=0$ channel is sufficiently attractive to generate a quasi-bound state, the $\Lambda(1405)$ resonance, below the $\bar{K}N$ threshold. Based on this picture, it may be expected that the $\bar{K}N$ interaction also generates quasi-bound states in kaon-nuclear systems, sometimes called kaonic nuclei.  At the same time, the $\bar{K}N$ quasi-bound picture of the $\Lambda(1405)$ is also related to the discussion of hadronic molecules in hadron spectroscopy.
Here an overview is presented of the theoretical studies developed for kaon-nucleon and kaon-nuclear systems. We start from the modern understanding of the $\Lambda(1405)$ resonance. We then discuss the $\bar{K}N$ interaction and various aspects of few-body kaonic nuclei. Heavier kaon-nuclear systems are examined from the viewpoint of nuclear many-body physics, with focus on the properties of antikaons in nuclear matter. Related topics, such as the $K^{-}p$ momentum correlation functions in high-energy collisions and the studies of kaonic atoms, are also discussed. 
}

\section{\textit{Introduction}}

\begin{figure}[tbp]
  \centering
  \includegraphics[height=50mm,angle=-00
  ]{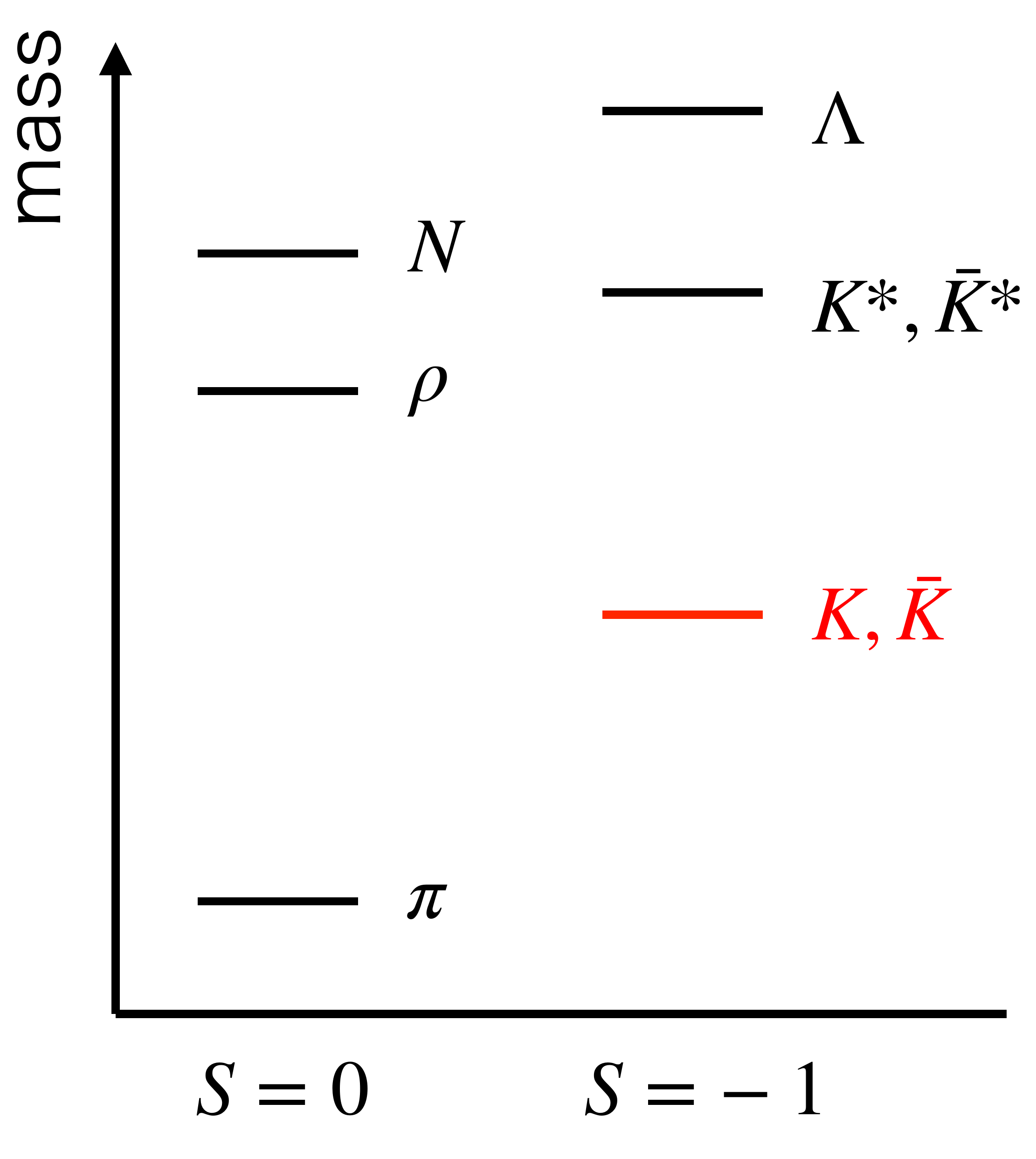}
  \caption{Masses of the lowest-lying hadrons in the strangeness $S=0$ and $S=-1$ sectors.}
  \label{fig:masses}
\end{figure}

The kaons ($K=K^{+},K^{0}$) and antikaons ($\bar{K}=K^{-},\bar{K}^{0}$) are the lightest pseudoscalar mesons with strangeness~\cite{ParticleDataGroup:2020ssz}. Their nature is closely related to the symmetry breaking pattern in low-energy QCD. In the $SU(3)$ octet of the light pseudoscalar mesons, kaons figure as flavor partners of the pions, the Nambu-Goldstone (NG) bosons associated with the spontaneous breaking of chiral symmetry in QCD. While the smallness of the pion mass has its origin in the almost vanishing up- and down-quark masses ($m_u \simeq 2$ MeV, $m_d \simeq 5$ MeV at a renormalization scale of 2 GeV), kaons are relatively massive because of the more sizable strange quark mass ($m_s \simeq 0.1$ GeV) reflecting the pronounced explicit chiral symmetry breaking in the strangeness sector. These features are illustrated in Fig.~\ref{fig:masses} where the masses of the lowest lying hadrons are plotted:  the kaons are not as light as the pions, but at the same time not as heavy as the ordinary hadrons other than the NG bosons. This intermediate nature of the kaons leads to various interesting phenomena of nonperturbative QCD at low energies. 

One such aspect is the possible formation of bound states of nuclei with antikaons ($\bar{K}$). Because the $\bar{K}$ has isospin $I=1/2$, there are two independent components of the $\bar{K}N$ interaction, $I=0$ and $I=1$, as in the nuclear force. Moreover, the $NN$ and $\bar{K}N$ interactions share a common feature. The $NN$ interaction in the $I=0$ ($^{3}S_{1}$) channel is sufficiently attractive to form the deuteron  bound state, and the $I=1$ ($^{1}S_{0}$) channel is also known to have an attractive scattering length. In a similar way, despite the otherwise obviously different physics of the $NN$ and $\bar{K}N$ systems, the $\bar{K}N$ interaction in the $I=0$ channel also has a quasi-bound state below the threshold, the $\Lambda(1405)$, and the $I=1$ interaction is likewise considered to be attractive (see Table~\ref{tbl:interactions}). A qualitative difference, though, is that the deuteron is stable, whereas the $\bar{K}N$ system couples strongly to the lower-energy $\pi\Sigma$ and $\pi\Lambda$ channels.  Consequently, the $\Lambda(1405)$ as a quasibound state imbedded in the $\pi\Sigma$ continuum has a short lifetime, i.e. a large decay width, $\Gamma \simeq 50$ MeV. Nonetheless the question was raised whether a self-bound system can be obtained by replacing one of the nucleons in nuclei by an antikaon. In fact, shortly after the prediction~\cite{Dalitz:1959dn,Dalitz:1960du} and the experimental observation of the  $\Lambda(1405)$~\cite{Alston:1961zzd}, possible bound kaon-nuclear systems have been discussed~\cite{PL7.288}. Recent interest in the kaon-nuclear systems was triggered by the work of Akaishi and Yamazaki~\cite{Akaishi:2002bg} who proposed the existence of possible deeply bound kaonic nuclei with a narrow width. Since then, many theoretical and experimental investigations have been devoted kaon-nuclear systems with these issues in mind. 

\begin{table}[btp]
\caption{Qualitative features of the $NN$ and $\bar{K}N$ $s$-wave interactions.}
\begin{center}

\begin{tabular}{|l|l|l|}
\hline
 & $I=0$ & $I=1$ \\ \hline
  $NN$ 
  & $d$ (bound state)
  & attractive \\
  $\bar{K}N$
  & $\Lambda(1405)$ (quasi-bound state)
  & attractive \\
\hline
\end{tabular}
\end{center}
\label{tbl:interactions}
\end{table}%

This chapter introduces the theoretical  framework for kaon-nuclear systems, starting with the $\Lambda(1405)$ as a prototype example. We then describe few-body and many-body kaon-nuclear systems, including a discussion of kaon properties in nuclear matter. We also refer here to several review articles covering related topics~\cite{Ramos2001,Friedman:2007zza,Hyodo:2011ur,Gal:2016boi,Tolos:2020aln,Meissner:2020khl,Mai:2020ltx,Hyodo:2020czb}.

\section{\textit{The $\Lambda(1405)$ resonance}}

The $\Lambda(1405)$ resonance is nominally located just 27 MeV below the $K^-p$ threshold ($M_p + m_{K^-} = 1432$ MeV). Its properties are therefore closely linked to the low-energy $\bar{K}N$ interaction. The $\Lambda(1405)$ is the lowest lying excited baryon in the strangeness $S=-1$ and isospin $I=0$ sector, rated as a four-star resonance in PDG~\cite{ParticleDataGroup:2020ssz}. Studies of the $\Lambda(1405)$ go back to 1959 when Dalitz and Tuan theoretically predicted a resonance below the $\bar{K}N$ threshold~\cite{Dalitz:1959dn,Dalitz:1960du}. Based on the coupled-channels unitarity of the $\bar{K}N$-$\pi\Sigma$ system and the empirical $\bar{K}N$ scattering length, they deduced the existence of a resonance in the $\pi\Sigma$ scattering amplitude. Shortly after the prediction, experimental evidence for the $\Lambda(1405)$ was reported in the $\pi\Sigma$ invariant mass distribution of the $K^-p\to \pi\pi\pi\Sigma$ reaction at 1.15 GeV~\cite{Alston:1961zzd}. Since then a large amount of theoretical and experimental investigations have been devoted to clarify the nature of the $\Lambda(1405)$ (see Refs.~\cite{Hyodo:2011ur,Meissner:2020khl,Mai:2020ltx,Hyodo:2020czb}). In particular, recent developments have firmly established the following basic properties of the $\Lambda(1405)$:
\begin{itemize}

\item Eigenenergy: The generalized eigenenergy of an unstable state is expressed by the position of the resonance pole of the scattering amplitude in the complex energy plane (the real and imaginary parts respectively correspond to the mass and half width of the resonance). The pole positions of the $\Lambda(1405)$ have been pinned down~\cite{Ikeda:2011pi,Ikeda:2012au,Guo:2012vv,Mai:2014xna}, thanks to accurate constraints from the precise measurement of kaonic hydrogen by the SIDDHARTA collaboration~\cite{SIDDHARTA:2011dsy,Bazzi:2012eq}.

\item Spin and parity: Experimental determination of the spin and parity has been carried out by the CLAS collaboration in the photoproduction $\gamma p\to K^{+}\Lambda(1405)$~\cite{CLAS:2014tbc}. The result confirms the expected quantum numbers of $J^{P}=1/2^{-}$.
\end{itemize}

Current interest on the $\Lambda(1405)$ is focused on its internal structure. In conventional constituent quark models, the negative parity excited baryons are described by the internal excitation of three quarks in the confining potential. In a systematic study of negative parity baryons using such a quark model~\cite{Isgur:1978xj}, it is shown that the mass of the $\Lambda(1405)$ deviates significantly (by $\sim 100$ MeV) from the standard quark model prediction, in contrast to other negative parity states which are well described by the quark model. This suggests that the $\Lambda(1405)$ has a  more exotic internal structure, unlike that of a simple three-quark state. In particular, the static quark model picture lacks the dynamical aspect of the excited baryons which couple to the meson-baryon continuum channels. An alternative picture of the $\Lambda(1405)$, as a weakly bound $\bar{K}N$ molecule induced by the attractive $\bar{K}N$ interaction, is in line with Ref.~\cite{Dalitz:1967fp}. This picture is also supported by a recent analysis of the  compositeness~\cite{Kamiya:2015aea,Kamiya:2016oao} and emphasizes once again the attractive nature of the $\bar{K}N$ interaction. In this sense, the internal structure of the $\Lambda(1405)$ is important, not only in hadron spectroscopy to search for non-conventional configurations of hadrons, but also for the deeper understanding of the $\bar{K}N$ interaction and its implications for kaon-nuclear physics. 

As a resonance in $\pi\Sigma$ scattering located close to the $\bar{K}N$ threshold, the $\Lambda(1405)$ requires a dynamical treatment in terms of coupled-channel meson-baryon scattering. The pseudoscalar $\pi$ and $\bar{K}$ are combined with the respective $1/2^{+}$ baryons in $s$ wave to form the $J^{P}=1/2^{-}$ baryonic system with strangeness $S=-1$. The lowest lying pseudoscalar mesons are regarded as Nambu-Goldstone bosons associated with the spontaneous breaking of $N_{f}=3$ chiral symmetry. Therefore their low-energy $s$-wave interactions with the octet baryons are constrained by the Weinberg-Tomozawa theorem~\cite{Weinberg:1966kf,Tomozawa:1966jm}. Based on this observation, a series of works~\cite{Kaiser:1995eg,Oset:1997it,Oller:2000fj} developed a successful theoretical framework, called chiral $SU(3)$ dynamics, in which the coupled-channel meson-baryon scattering amplitude satisfying the unitarity condition is constructed in a systematic manner. 

\begin{figure}[tbp]
  \centering
  \includegraphics[width=10cm,bb=0 0 1137 143]{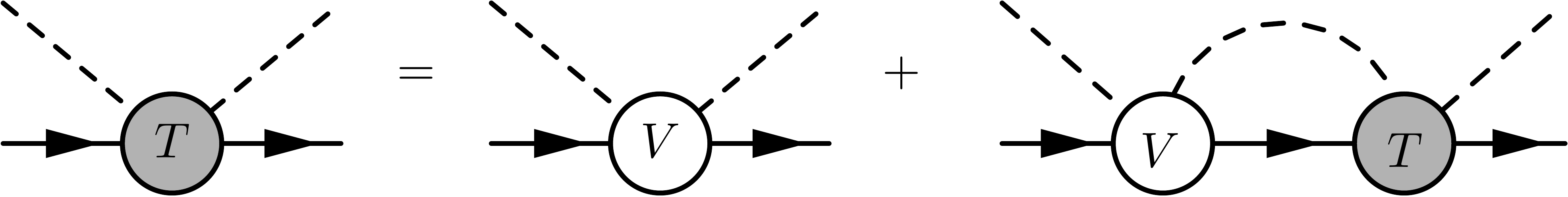}
  \caption{Feynman diagrams for the scattering equation. Solid (dashed) lines represent the baryon (meson) propagator. Open (shaded) blobs correspond to the interaction kernel $V$ (scattering amplitude $T$) and the meson-baryon intermediate loop stands for the loop function $G$. }
  \label{fig:scatteringEq}
\end{figure}

Denoting the meson-baryon channels by the indices $i,j$, the scattering amplitude takes a matrix form, $T_{ij}$, in channel space. In the s-wave ($\ell=0$) channels, $T_{ij}$ is a function of the total meson-baryon energy $W$. In chiral $SU(3)$ dynamics, $T_{ij}$ is then given by the solution of a coupled-channel Lippmann-Schwinger scattering equation;
\begin{align}
   T_{ij}
   =V_{ij}+V_{ik}G_{k}T_{kj} ,
   \label{eq:scattering}
\end{align}
where the summation over the repeated indices is implicit. In Fig.~\ref{fig:scatteringEq}, Feynman diagram representation of the scattering equation is shown. The kernel $V_{ij}$ represents the meson-baryon interaction,  constructed so that it systematically satisfies the chiral symmetry constraints ~\cite{Ecker:1994gg,Bernard:1995dp,Pich:1995bw,Bernard:2007zu,Scherer:2012xha}. Furthermore $G_{k}$ is the loop function representing the propagation of the meson and baryon in the intermediate state with index $k$.  Substituting $T_{kj}$ on the right-hand-side of Eq.~\eqref{eq:scattering} iteratively, one finds
\begin{align}
   T_{ij}
   =V_{ij}+V_{ik}G_{k}V_{kj} +V_{ik}G_{k}V_{kl}G_{l}V_{lj}+\dotsb,
   \label{eq:scatteringitterative}
\end{align}
where the multiple scatterings in coupled channels are iterated to all orders in $T_{ij}$. This guarantees the unitarity of the scattering amplitude. The same strategy is applied in the study of the nuclear force within the framework of chiral effective field theory~\cite{Epelbaum:2008ga,Machleidt:2011zz}. In the following, we present the properties of $V_{ij}$ and $G_{i}$.

Chiral perturbation theory~\cite{Ecker:1994gg,Bernard:1995dp,Pich:1995bw,Bernard:2007zu,Scherer:2012xha} classifies the interaction kernel $V$ by the chiral order in terms of a low-momentum scale $p$. The chiral order, denoted by $\mathcal{O}(p^{n})$, refers to an expansion in powers of $p/\Lambda_{\chi}$ where $\Lambda_{\chi}\sim 1$ GeV is the characteristic scale of the spontaneous breaking of chiral symmetry. The terms with smaller chiral orders dominate  in the low-energy region, $p\ll \Lambda_{\chi}$, thus low-energy phenomena can be described by a finite number of terms.

In meson-baryon scattering the leading order (LO) terms start at  $\mathcal{O}(p)$ and involve couplings of the $SU(3)$ pseudoscalar meson and $J^{P}=1/2^+$ baryon octets through their vector and axial vector currents. For example, the dominant Weinberg-Tomozawa (WT) term in the $K^-$ proton elastic scattering channel derives from the following piece of the chiral interaction Hamiltonian:
\begin{align}
   \delta H_{\rm WT}(K^-p\rightarrow K^-p) = 
 -{i\over 2f^2}\int d^3x\,\bar{\psi}_p(x)\gamma^\mu\psi_p(x)\,K^+(x)\partial_\mu K^-(x)~~,
\label{eq:HWT}
\end{align}
where $\psi_p$ and $K^-$ are the proton and antikaon fields, respectively. Note that the WT term is completely determined by the pseudoscalar meson decay constant, $f \simeq 0.1$ GeV. The interaction vanishes in the limit of zero meson four-momentum, $q^\mu = (E,\vec{q})$, a feature characteristic of Nambu-Goldstone bosons. The s-wave interaction is governed by the time component of the vector current. In the limit of a pointlike, static nucleon, the leading WT term is therefore a contact interaction proportional to $E/f^2$, with the meson energy $E= \sqrt{m^2 + \vec{q}^2}$. At threshold the WT interaction kernels in $K^-p$ and $K^-n$ channels have the form
\begin{align}
V_{\rm WT}(K^-p) = 2 \,V_{\rm WT}(K^-n) \propto - {m_K\over f^2}~~.
\label{eq:VWT}
\end{align}
For $K^+p$ and $K^+n$ scattering the corresponding terms are repulsive, of equal magnitude as in Eq.~\eqref{eq:VWT} but with opposite sign. The important point to note here is that the leading $\bar{K}N$ interaction close to threshold is much stronger than the s-wave $\pi N$ interaction. It scales with the kaon mass $m_K$ (rather than the small pion mass $m_\pi$) and reflects the stronger explicit chiral symmetry breaking characteristic of the strange quark. 

\begin{figure}[tbp]
  \centering
  \includegraphics[width=11cm,bb=0 0 1533 143]{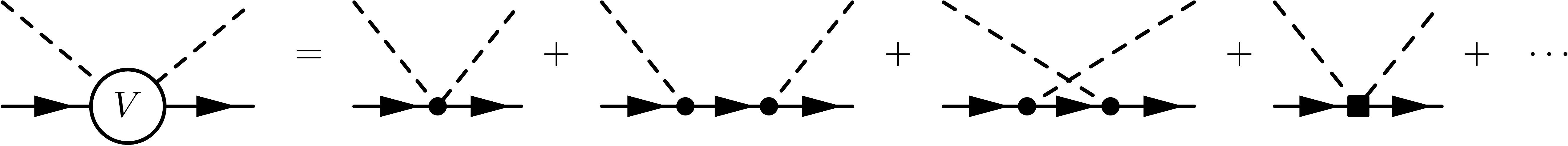}
  \caption{Feynman diagrams for the interaction kernel $V$. Black dots represent the $\mathcal{O}(p)$ vertices, and black square stands for the $\mathcal{O}(p^{2})$ vertex. The first term in the right hand side is $V_{\rm WT}$, the second and third terms are $V_{\rm Born}$, the forth term is $V_{\rm NLO}$, and ellipsis shows the $\mathcal{O}(p^{3})$ contributions.}
  \label{fig:V}
\end{figure}

The meson-baryon interaction up to and including the next-to-leading order (NLO) terms of $\mathcal{O}(p^{2})$ is schematically written as
\begin{align}
   V
   =\underbrace{V_{\rm WT}
   +V_{\rm Born}}_{\mathcal{O}(p^{1})}
   +\underbrace{V_{\rm NLO}}_{\mathcal{O}(p^{2})}
   +\dotsb ~~.
\end{align}
The corresponding Feynman diagrams are shown in Fig.~\ref{fig:V}. The ellipsis stands for the $\mathcal{O}(p^{3})$ and higher order contributions. In each interaction term there are low-energy constants (LECs), whose strength cannot be determined from symmetry arguments alone. The precision of the calculation can be increased by introducing higher order terms, but this requires a sufficient amount of experimental data to determine all the LECs. 

The dominant contribution in the leading order $\mathcal{O}(p^{1})$ terms is the Weinberg-Tomozawa term $V_{\rm WT}$ (Fig.~\ref{fig:V}, first term in the right hand side). In the chiral dynamics framework the pertinent low-energy theorems are automatically built in, and the low-energy limit of $V$ reduces to $V_{\rm WT}$ satisfying the Weinberg-Tomozawa theorem. As mentioned, $V_{\rm WT}$ has no LEC apart from the pseudoscalar meson decay constant. The sign and strength of the interaction are determined by flavor $SU(3)$ symmetry.

The Born terms $V_{\rm Born}$ are constructed by the $s$- and $u$-channel baryon exchange diagrams (Fig.~\ref{fig:V}, second and third terms). The strength parameters of the meson-baryon-baryon three-point vertices in these diagrams are given by the axial vector coupling constants of the participating baryons, satisfying Goldberger-Treiman relations~\cite{Goldberger:1958vp}. Their values reflect the internal structure of the baryons. While $V_{\rm Born}$ is counted as $\mathcal{O}(p)$, the main contribution of this term is in $p$-wave, and the $s$-wave projected part appears at higher order than $V_{\rm WT}$ in the non-relativistic expansion~\cite{Weinberg:1996kr}. Thus, in the low-energy limit, $V_{\rm Born}$ is much smaller than $V_{\rm WT}$, and the meson-baryon interaction is model-independently given by the leading Weinberg-Tomozawa term, thanks to chiral symmetry. When confronted with high-precision data, such as the measurement of kaonic hydrogen by the SIDDHARTA collaboration~\cite{SIDDHARTA:2011dsy,Bazzi:2012eq}, the precision of the theoretical framework is increased by introducing the higher order terms. This is achieved by including the NLO terms $V_{\rm NLO}$ at $\mathcal{O}(p^{2})$~\cite{Ikeda:2011pi,Ikeda:2012au,Guo:2012vv,Mai:2014xna} (fourth terms in Fig.~\ref{fig:V}).

The scattering equation \eqref{eq:scattering} is in general an integral equation, with the off-shell structure of the interaction kernel $V_{ij}$ being integrated over. In practical applications the equation is usually reduced into  algebraic form by using the on-shell factorization. This formulation preserves the unitarity condition. It is also  justified by the N/D method (see Refs.~\cite{Oset:1997it,Oller:2000fj,Hyodo:2011ur,Mai:2020ltx}). Because the leading order $V_{\rm WT}$ is a contact interaction,  the momentum integration in the loop function involves an  ultraviolet divergence. This divergence is usually tamed by dimensional regularization, and the finite part of the loop function $G_{i}$ is specified by a subtraction constant  which plays the role of an ultraviolet cutoff. The one-loop diagram is counted as $\mathcal{O}(p^3)$, and therefore the renormalization procedure of the meson-baryon scattering is completed at $\mathcal{O}(p^3)$. In the unitarized amplitude in Eq.~\eqref{eq:scattering} with the $\mathcal{O}(p^2)$ interaction kernel ($V=V_{\rm WT}+V_{\rm Born}+V_{\rm NLO}$), the subtraction constants need to be determined by fitting experimental data. As mentioned, $V_{\rm WT}$ is completely determined by chiral symmetry. The Born terms contain the axial vector coupling constants of the octet baryons, conventionally denoted by $D$ and $F$, which are determined by the semi-leptonic decays of hyperons. Thus, when the $\mathcal{O}(p^{1})$ interactions $V_{\rm WT}$ and $V_{\rm Born}$ are used, the only free parameters are the subtraction constants. In the strangeness $S=-1$ coupled-channel scattering matrix there are six isospin states ($\bar{K}N,\pi\Lambda,\pi\Sigma, \eta\Lambda,\eta\Sigma,K\Xi$), and six subtraction constants can be used to fit the experimental data. In $V_{\rm NLO}$, there are seven LECs, which should be determined also by the experimental data in the $\mathcal{O}(p^{2})$ calculations, in addition to the subtraction constants. In the current situation the LECs up to $\mathcal{O}(p^{2})$ terms can be determined by experimental data, but in order to perform $\mathcal{O}(p^{3})$ calculations, the data base is not (yet) sufficient. 

\begin{figure}[tbp]
  \centering
  \includegraphics[height=65mm,angle=-00,bb=0 0 1920 1080]{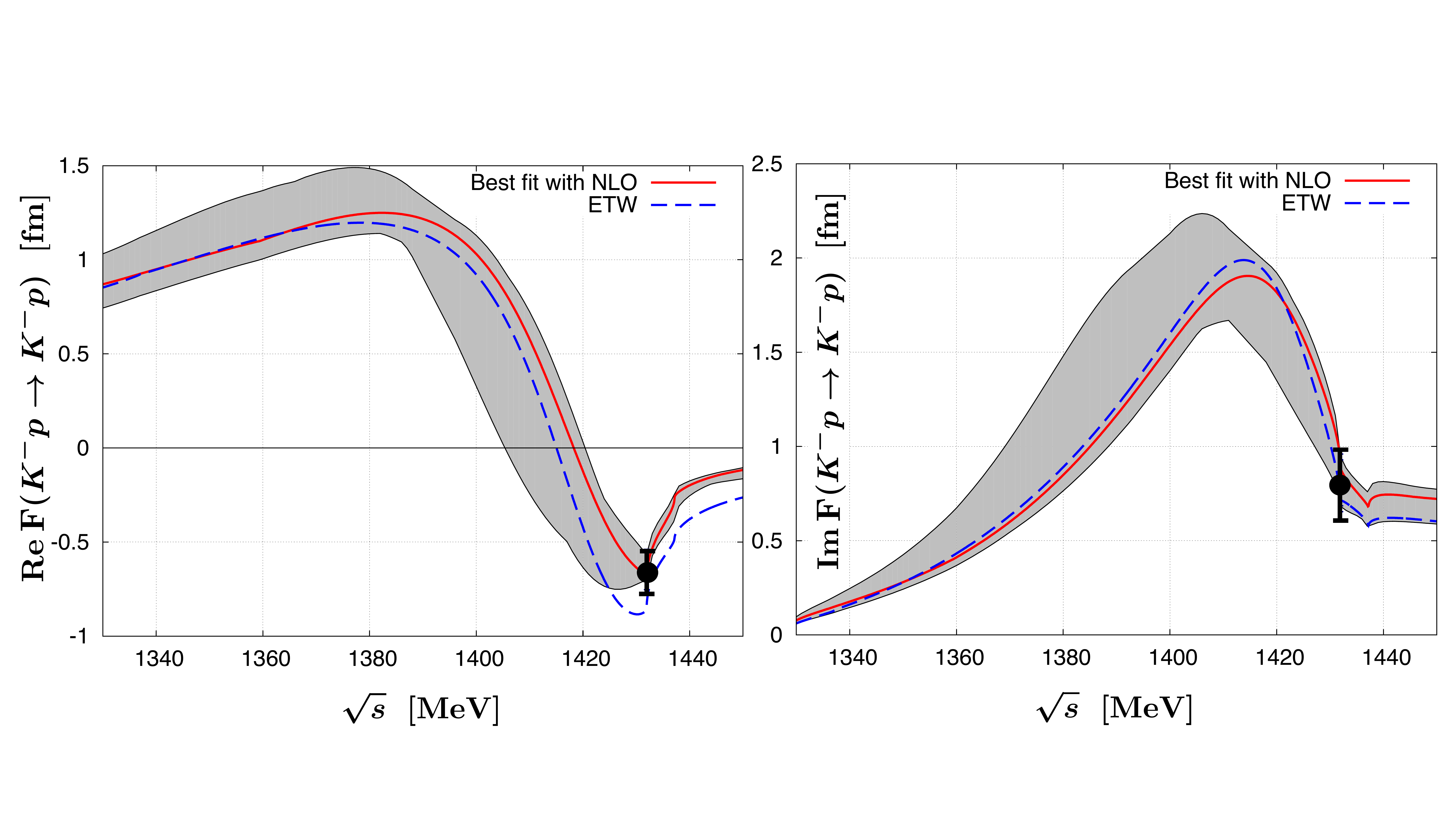}
  \caption{Real and imaginary parts of the $K^-p \rightarrow K^-p$ forward scattering amplitude, $F = T/8\pi\sqrt{s}$, from chiral $SU(3)$ dynamics at NLO (solid curve) and constrained by the SIDDHARTA kaonic hydrogen measurements. The dot at threshold marks the corresponding deduced scattering length. Uncertainties are represented by the grey band. Also shown (dashed curve) for comparison is an effective Weinberg-Tomozawa amplitude in a reduced three-channel ($\bar{K}N-\pi\Sigma-\pi\Lambda$) scheme using values $f_\pi = 92.4$ MeV, $f_K=109$ MeV of the meson decay constants and three subtraction constants as fit parameters. Adapted from Ref.~\cite{Ikeda:2012au}.}
  \label{fig:amplitude}
\end{figure}

Figure~\ref{fig:amplitude} shows the result of a chiral $SU(3)$ dynamics calculation of the $K^-p$ forward scattering amplitude~\cite{Ikeda:2012au}, featuring the prominent emergence of the $\Lambda(1405)$ as a $\bar{K}N$ quasibound state imbedded in the $\pi\Sigma$ continuum. This amplitude has served as input to many investigations of $\bar{K}$ interactions in more complex systems.

\begin{figure}[tbp]
  \centering
  \includegraphics[height=50mm,angle=-00
  ]{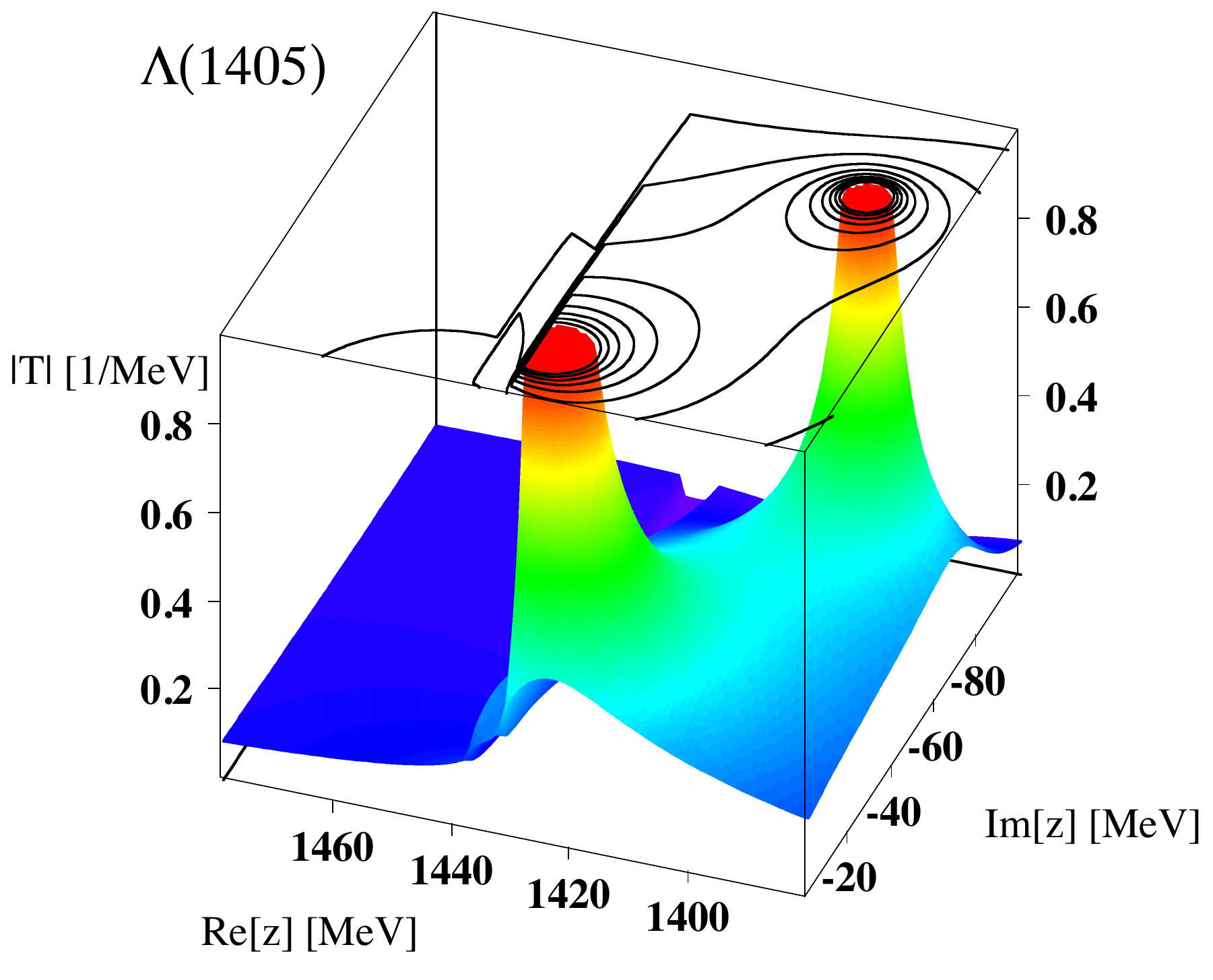}
  \caption{The absolute value of the scattering amplitude in the complex energy $z$ plane. Adapted from Ref.~\cite{Hyodo:2011ur}.}
  \label{fig:pole1405}
\end{figure}

Poles of the scattering amplitude $T_{ij}$ in the complex energy plane represent the generalized eigenenergies of  unstable resonances~\cite{Hyodo:2020czb}. Usually there is a single pole for a resonance state, whose position $z$ in the complex energy plane is related to the mass $M_{R}$ and the width $\Gamma_{R}$ as 
\begin{align}
   M_{R}
   =\text{Re } z,\quad
   \Gamma_{R}
   =-2\ \text{Im } z .
\end{align}
It was first pointed out in Ref.~\cite{Oller:2000fj} that there are actually two poles with $I=0$ and $S=-1$ in the $\Lambda(1405)$ energy region. This fact has been confirmed in many subsequent studies~\cite{Meissner:2020khl,Mai:2020ltx,Hyodo:2020czb}, including the cloudy bag model~\cite{Fink:1989uk}, meson-exchange model~\cite{Haidenbauer:2010ch}, dynamical coupled-channel model~\cite{Kamano:2014zba,Kamano:2015hxa}, and Hamiltonian effective field theory~\cite{Liu:2016wxq}. One pole is located near the $\bar{K}N$ threshold with a narrow decay width. A second pole is found around the $\pi\Sigma$ threshold with relatively broad width. Namely, the nominal ``$\Lambda(1405)$'' is not a single resonance. It is realized as a superposition of two eigenstates in the same energy region. In fact, in the latest version of PDG~\cite{ParticleDataGroup:2020ssz}, the lower energy pole is called $\Lambda(1380)$ as a new two-star resonance, and $\Lambda(1405)$ is reserved for the higher energy pole around 1420 MeV. Here we note that the two resonance poles in the complex energy plane do not necessarily exhibit two peaks in the projected spectrum on the real axis. The shape of the spectrum depends on the relative phase of the residues of the poles. In the $\Lambda(1405)$ case there is only one broad and asymmetric peak structure on the real axis (see Fig.~\ref{fig:pole1405}). In other words, the $\Lambda(1405)$ spectrum is produced by the contributions from both poles. 

The origin of the two poles can be understood by the properties of the Weinberg-Tomozawa interaction~\cite{Jido:2003cb,Hyodo:2007jq}. The meson-baryon channel can be expressed either in the physical basis (such as $K^{-}p$, $\bar{K}^{0}n$, \dots), in the isospin basis ($\bar{K}N(I=0)$, $\bar{K}N(I=1)$, \dots), or in the $SU(3)$ basis ($\bm{1}(I=0)$, $\bm{8}(I=0)$, \dots). The basis transformation can be done through the $SU(2)$ Clebsch-Gordan coefficients and the $SU(3)$ isoscalar factors.  Representing the interaction in the isospin basis and in the $SU(3)$ basis is useful in order to study the symmetry underlying the interaction. The coupling strength of $V_{\rm WT}$ is $SU(3)$ symmetric, hence $V_{\rm WT}$ becomes a diagonal matrix in the $SU(3)$ basis where the mixing of different representations is absent. Therefore, in the $SU(3)$ symmetric limit of hadron masses, the coupled-channel equation reduces to a set of independent single-channel problems~\cite{Hyodo:2006yk,Hyodo:2006kg}. There are four channels with $I=0$ ($\bm{1}$, $\bm{8}$, $\bm{8}^{\prime}$, and $\bm{27}$). Three of those are shown to be sufficiently attractive to generate bound states in the $SU(3)$ limit~\cite{Jido:2003cb}.  It is found that one of the poles evolves into the $\Lambda(1670)$ resonance. The other two poles move towards the $\Lambda(1405)$ energy region, along with the $SU(3)$ breaking towards physical masses.  In the same way,  there are four channels in the isospin basis ($\pi\Sigma$, $\bar{K}N$, $\eta\Lambda$, and $K\Xi$), and $V_{\rm WT}$ is attractive in the diagonal $\pi\Sigma$, $\bar{K}N$ and $K\Xi$ channels. In the absence of off-diagonal couplings, the $\bar{K}N$ attraction generates a bound state below the $\bar{K}N$ threshold, and the $\pi\Sigma$ attraction creates a resonance above the $\pi\Sigma$ threshold~\cite{Hyodo:2007jq}. By introducing the channel couplings gradually, these eigenstates evolve into the two poles between the $\bar{K}N$ and $\pi\Sigma$ thresholds. In this way, the origin of the two poles can be understood by the two attractive components in the Weinberg-Tomozawa term, which are model-independently constrained by the low-energy theorem of chiral symmetry.

Now we explain how the pole positions in PDG are determined, based on Refs.~\cite{Ikeda:2011pi,Ikeda:2012au,Guo:2012vv,Mai:2014xna}. In these works of NLO chiral $SU(3)$ dynamics, a systematic uncertainty analysis was performed with inclusion of the SIDDHARTA kaonic hydrogen measurement. The experimental data base for this analysis can be classified as follows:\footnote{The $K^{-}p$ correlation function data was not available when Refs.~\cite{Ikeda:2011pi,Ikeda:2012au,Guo:2012vv,Mai:2014xna} were published, but the data in Ref.~\cite{ALICE:2019gcn} is shown to be consistent with the scattering amplitude of Ref.~\cite{Ikeda:2011pi,Ikeda:2012au}, as we describe below~\cite{Kamiya:2019uiw}.}
\begin{itemize}
\item[(i)]\ \ elastic and inelastic total cross sections of $K^{-}p$ scattering,
\item[(ii)]\ \ threshold branching ratios~\cite{Tovee:1971ga,Nowak:1978au},
\item[(iii)]\ \ energy level shift and width of the kaonic hydrogen~\cite{SIDDHARTA:2011dsy,Bazzi:2012eq},
\item[(iv)]\ \ $\pi\Sigma$ invariant mass distributions in various reactions~\cite{Ahn:2003mv,Niiyama:2008rt,CrystallBall:2004ovf,Zychor:2007gf,HADES:2012csk,CLAS:2013rjt,CLAS:2013zie}.
\end{itemize}
The total cross section $\sigma_{ij}$ from channel $j$ to $i$ at the energy $W= \sqrt{s}$ is derived from $T_{ij}$ as (here we follow the convention in Refs.~\cite{Ikeda:2011pi,Ikeda:2012au}):
\begin{align}
   \sigma_{ij}(W)
   =\frac{q_{i}}{q_{j}}\frac{|T_{ij}(W)|^{2}}{16\pi W^{2}} ,
   \label{eq:crosssection}
\end{align}
where $q_{i}$ is the magnitude of the three-momentum in channel $i$. The threshold branching ratios are calculated by the combination of the cross section at the $K^{-}p$ threshold, $\sigma_{ij}(W=m_{K^{-}}+M_{p})$. The energy shift $\Delta E$ and width $\Gamma$ of kaonic hydrogen are related to the $K^{-}p$ scattering length $a_{K^{-}p}$ through the improved Deser formula~\cite{Meissner:2004jr}, and $a_{K^{-}p}$ can be calculated from the scattering amplitude as
\begin{align}
   a_{K^{-}p}
   =\left.\frac{T_{K^{-}p,K^{-}p}(W)}{8\pi W}\right|_{W=m_{K^{-}}+M_{p}} .
\end{align}
In this way, data sets (i)-(iii) are related to the two-body scattering amplitude $T_{ij}$ and can be used as direct constraints on $T_{ij}$. In contrast, the $\pi\Sigma$ spectra cannot be calculated in terms of the corresponding scattering amplitude $T_{ij}$ only. Elastic $\pi\Sigma$ scattering experiments are practically not possible, so the $\pi\Sigma$ spectra can only be measured in production reactions, where particles other than $\pi\Sigma$ exist in the final state. Consider for example the simplest case of a three-body final state, such as it appears in the photoproduction reaction,  $\gamma p\to K^{+}\pi\Sigma$. Let channel $i$ denote the $\pi\Sigma$ subsystem in the final state. The distribution of the invariant $\pi\Sigma$ mass $M_{I}$ can be written as 
\begin{align}
   \frac{d\sigma_{i}(M_{I})}{dM_{I}}
   \propto \left|\sum_{j}T_{ij}(M_{I})G_{j}(M_{I})C_{j}\right|^{2} ~~.
   \label{eq:Minv}
\end{align}
The coefficients $C_{j}$ represent the weight of the initial channel $j$, which reflect various aspects of the reaction, such as the kinematics (initial energy, scattering angle of $K^{+}$, etc.). In order to extract the information of the two-body scattering amplitude $T_{ij}$, one needs to estimate the $C_{j}$s in some way. In addition, Eq.\,\eqref{eq:Minv} does not include the final state interactions of other hadron pairs ($\pi K$ and $K\Sigma$), so the full three-body dynamics is missing. Therefore, in comparison with the direct constraints (i)-(iii),  using the $\pi\Sigma$ spectra (iv) requires special care.  

\begin{figure}[tbp]
  \centering
  \includegraphics[height=45mm,angle=-00,bb=0 0 1920 1080]{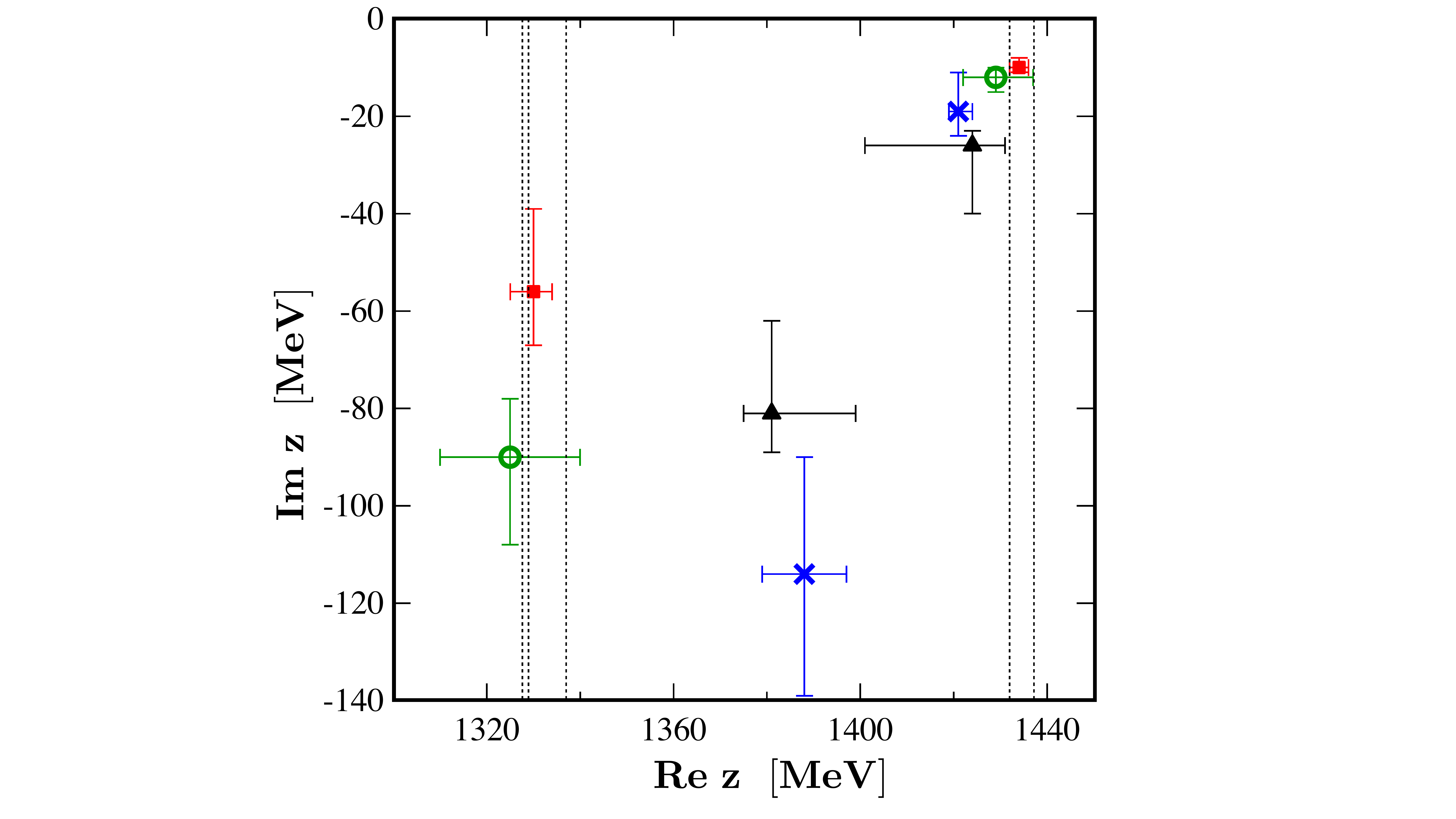}
  \caption{Resonance poles of $\Lambda(1405)$ and $\Lambda(1380)$ in the complex energy ($z$) plane. Triangles, squares, crosses, and circles show the results in Refs.~\cite{Ikeda:2011pi,Ikeda:2012au} (NLO), Ref.~\cite{Guo:2012vv} (Fit II), Ref.~\cite{Mai:2014xna} (Solution \#2), and Ref.~\cite{Mai:2014xna} (Solution \#4). Dotted lines represent the threshold energies of meson-baryon channels: $\pi^{0}\Sigma^{0}$,  $\pi^{-}\Sigma^{+}$,  $\pi^{+}\Sigma^{-}$,  $K^{-}p$,  $\bar{K}^{0}n$ from left to right. Adapted from Ref.~\cite{Hyodo:2020czb}.}
 \label{fig:poles}
	
\end{figure}

In Refs.~\cite{Ikeda:2011pi,Ikeda:2012au}, the meson-baryon scattering amplitude was constructed using NLO chiral $SU(3)$ dynamics together with the constraints (i)-(iii). A systematic uncertainty analysis was performed to determine the subtraction constants and the LECs in the NLO term, achieving an accuracy $\chi^{2}/{\rm d.o.f}=0.96$. This indicates that the ``puzzle'' caused by the DEAR experiment~\cite{DEAR:2005fdl}, a possible inconsistency of the scattering data and an earlier kaonic hydrogen measurement~\cite{Borasoy:2004kk,Borasoy:2005ie}, has been resolved. Even when the interaction is restricted to $V_{\rm WT}$ only, the overall description is reasonable ($\chi^{2}/{\rm d.o.f}=1.12$). This means that the scattering amplitude is essentially determined by $V_{\rm WT}$, while NLO corrections are required as a further improvement to deal with the accurate SIDDHARTA data. It also turns out that the SIDDHARTA result reduces the uncertainty of the subthreshold extrapolation of the scattering amplitude, in comparison to a similar analysis that only uses the data (i) and (ii)~\cite{Borasoy:2006sr}. In Ref.~\cite{Guo:2012vv}, in addition to (i)-(iii), the cross sections of $K^{-}p\to \eta\Lambda$~\cite{CrystalBall:2001uhc}, $\pi\Lambda$ phase shift at $\Xi^{-}$ mass~\cite{FNALE756:2003kkj,HyperCP:2004not}, and the $\pi\Sigma$ mass distributions in the $\Sigma^{+}(1660)\to \pi^{+}\pi^{-}\Sigma^{+}$ reaction~\cite{Hemingway:1984pz} and in the $K^{-}p\to \pi^{0}\pi^{0}\Sigma^{0}$ reaction~\cite{CrystallBall:2004ovf} were included in the analysis. The effect of the $SU(3)$ breaking in the meson decay constants was discussed in detail. In Ref.~\cite{Mai:2014xna}, the $\pi\Sigma$ spectra in the photoproduction data by CLAS~\cite{CLAS:2013rjt} was used in addition to (i)-(iii). It was shown that some solutions allowed by (i)-(iii) were rejected by the CLAS data, and two solutions were finally obtained. The results of the pole positions in Refs.~\cite{Ikeda:2011pi,Ikeda:2012au,Guo:2012vv,Mai:2014xna} are shown in Table~\ref{tbl:poles}, and plotted in Fig.~\ref{fig:poles}. In all those analyses, two poles are found in the energy region of the coupled $\bar{K}N$-$\pi\Sigma$ systems. The position of the $\Lambda(1405)$ pole (near the $\bar{K}N$ threshold) has small ambiguity and converges to the region around 1420 MeV. This reflects the strong constraint from the SIDDHARTA data at the $K^{-}p$ threshold. On the other hand, the pole position of the $\Lambda(1380)$ (near the $\pi\Sigma$ threshold) shows sizable ambiguities in different analyses. It is therefore desirable to determine that pole position quantitatively in future studies. 

\begin{table}[tbp]
\caption{Resonance poles of $\Lambda(1405)$ and $\Lambda(1380)$ in PDG~\cite{ParticleDataGroup:2020ssz}.}
\begin{center}
\begin{tabular}{|l|l|l|}
\hline
 & $\Lambda(1405)$ [MeV] & $\Lambda(1380)$ [MeV] \\ \hline
Refs.~\cite{Ikeda:2011pi,Ikeda:2012au} NLO 
  & $1424^{+7}_{-23}- i 26^{+3}_{-14}$
  &  $1381^{+18}_{-6}- i 81^{+19}_{-8}$ \\
Ref.~\cite{Guo:2012vv} Fit II 
  & $1421^{+3}_{-2}- i 19^{+8}_{-5}$ 
  & $1388^{+9}_{-9}- i 114^{+24}_{-25}$ \\
Ref.~\cite{Mai:2014xna} solution \#2
  & $1434^{+2}_{-2} - i \, 10^{+2}_{-1}$ 
  & $1330^{+4~}_{-5~} - i \, 56^{+17}_{-11}$\\
Ref.~\cite{Mai:2014xna} solution \#4
  & $1429^{+8}_{-7} - i \, 12^{+2}_{-3}$ 
  & $1325^{+15}_{-15} - i \, 90^{+12}_{-18}$\\
\hline
\end{tabular}
\end{center}
\label{tbl:poles}
\end{table}%

\begin{figure}[tbp]
  \centering
  \includegraphics[height=40mm,angle=-00
  ]{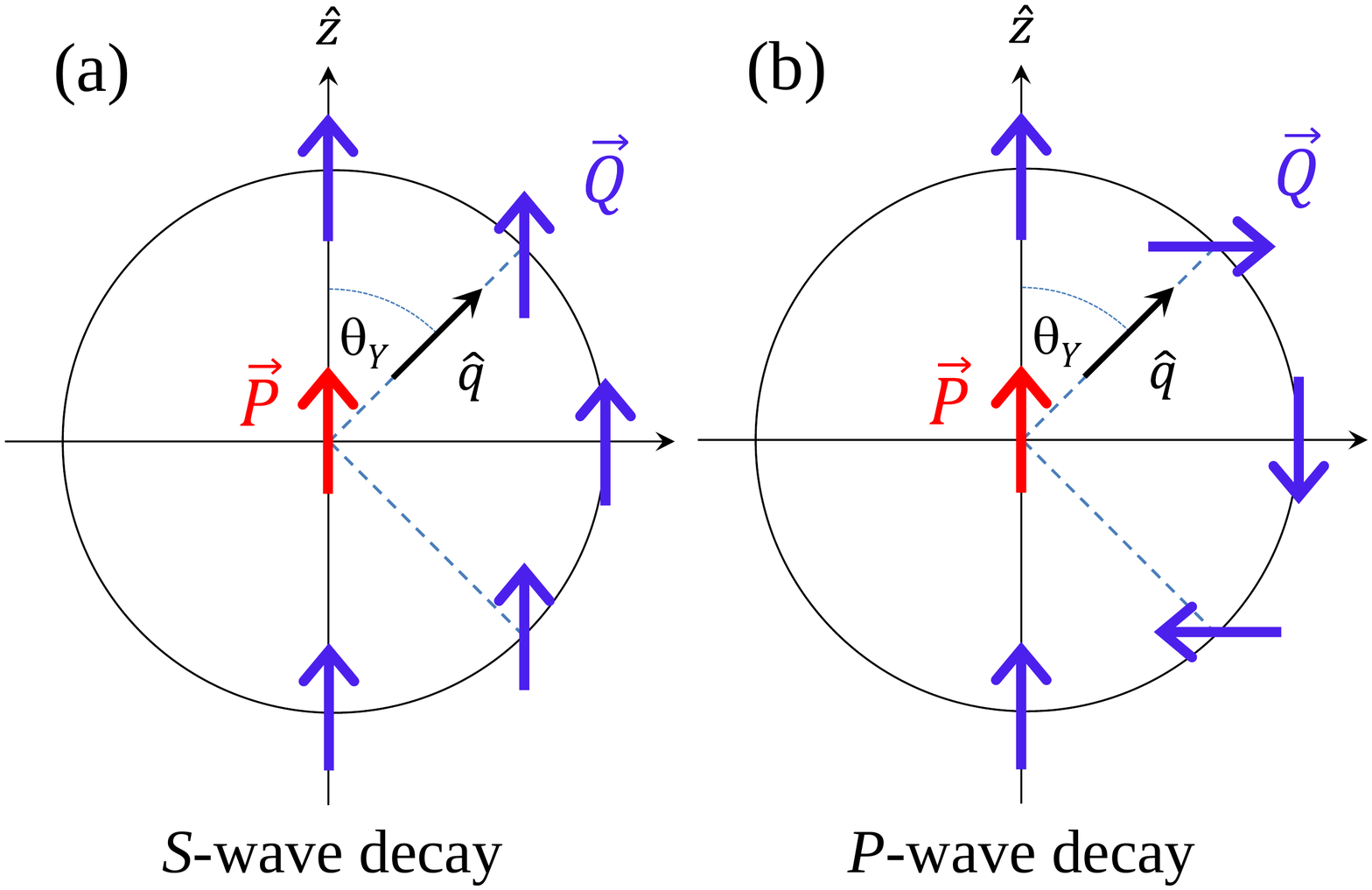}
  \includegraphics[height=35mm,angle=-00
  ]{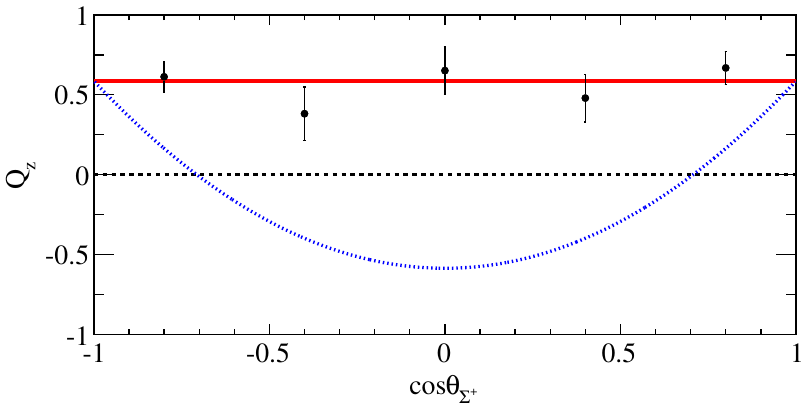}
  \caption{Upper figure: schematic illustration of the polarization transfer of $Y^* \rightarrow Y \pi$ decay with $Y^{*}$ having $J^{P}=1/2^{-}$ (a) or $J^{P}=1/2^{+}$ (b).  Lower figure: $z$-component of the $\Sigma^{+}$ polarization $Q_{z}$ as a function of the decay angle $\cos \theta_{\Sigma^{+}}$.  Solid line: average of the experimental data; dotted line: prediction for $p$-wave decay; dashed line marks the case without polarization. Adapted from Ref.~\cite{CLAS:2014tbc}.}
  \label{fig:CLAS_JP}
\end{figure}

The spin of the $\Lambda(1405)$ was considered to be consistent with $1/2$ in the past experiments~\cite{Engler:1965zz,Thomas:1973uh,Hemingway:1984pz}, but its parity had not been determined. In 2014, the direct experimental determination of the parity of the $\Lambda(1405)$ was performed in Ref.~\cite{CLAS:2014tbc} with the CLAS photoproduction data of $\gamma p\to K^{+}\Lambda(1405)$. Consider the decay angular distribution of the $\Lambda(1405)\rightarrow \pi^-\Sigma^+$ decay. In the unpolarized two-body decay via the strong interaction, the angular dependence is determined only by the spin, independent of the parity. If the polarization of the $\Sigma^+$ ($\vec{Q}$) and  that of the $\Lambda(1405)$ ($\vec{P}$) are specified, then the difference arises in the direction of $\vec{Q}$ with respect to $\vec{P}$, as shown in the left panel of Fig.~\ref{fig:CLAS_JP}. For instance, when the initial state is $J^{P}=1/2^{-}$ with an $s$-wave decay, the direction of $\vec{Q}$ is independent of the decay angle as in panel (a). On the other hand, for a $1/2^{+}$ state with a $p$-wave decay, the direction of $\vec{Q}$ rotates around the $\vec{P}$ vector as in panel (b). Thus, taking the $z$ axis in the direction of $\vec{P}$, the $z$ component of the $\Sigma$ polarization $Q_{z}$ is constant for $1/2^{-}$, while it changes sign depending on the decay angle for $1/2^{+}$. Experimentally, the $\Lambda(1405)$ polarization $\vec{P}$ is determined by the reaction plane specified by the photon and the $K^{+}$. The $\Sigma^{+}$ polarization $\vec{Q}$ is given by the weak decay of the $\Sigma^{+}$. The experimental result for the total energy $2.65<W<2.75$~GeV and the scattering angle of $K^{+}$ in the center-of-mass system, $0.70<\cos \theta_{K^{+}}^{\mathrm{cm} .}<0.80$, is shown in the right panel of Fig.~\ref{fig:CLAS_JP}. The observed $Q_{z}$ does not depend on the decay angle $\cos \theta_{\Sigma^{+}}$ with respect to $\vec{P}$, and hence the $J^{P}=1/2^{-}$ quantum numbers are confirmed for the $\Lambda(1405)$. 

\section{\textit{$K^{-}p$ correlation functions}}

In recent years, two-particle momentum correlation functions in the high-energy collisions have been of particular interest, as a new tool to extract properties of hadronic interactions that are not easily accessible in direct scattering experiments.  In high-energy heavy-ion and proton-proton collisions with high multiplicities in the TeV region,  hadrons are statistically emitted from the high-energy source created by the collisions~\cite{Bauer:1992ffu,Lisa:2005dd,ExHIC:2017smd}.  The produced hadrons can interact pairwise before they reach the detector. The momentum correlation of such a pair of hadrons reflects the effect of their two-body interaction.

 The correlation function of two hadrons (with momenta $\bm{p}_{1}$ and $\bm{p}_{2}$) is given by
\begin{align}
   C(\bm{q}) &= \frac{N_{12}(\bm{p}_{1},\bm{p}_{2})}{N_{1}(\bm{p}_{1})N_{2}(\bm{p}_{2})},
\end{align}
where $\bm{q}$ is the relative momentum of the hadron pair. The numerator represents the production yield with simultaneous detection of the hadron pair. The denominator corresponds to the product of the yields of the individually detected hadrons, with $N_{i}$ normalized to unity when integrated over the momentum. If the hadronic final state interactions were completely absent, the correlation function of a distinguishable hadron pair would be $C(\bm{q})=1$, and that of a pair of identical hadrons would be the correlation induced by quantum statistics.  Information about the hadron interactions can therefore be extracted by precise measurements of the deviation from these baseline correlations. 

\begin{figure}[tbp]
  \centering
  \includegraphics[height=48mm,angle=-00
  ]{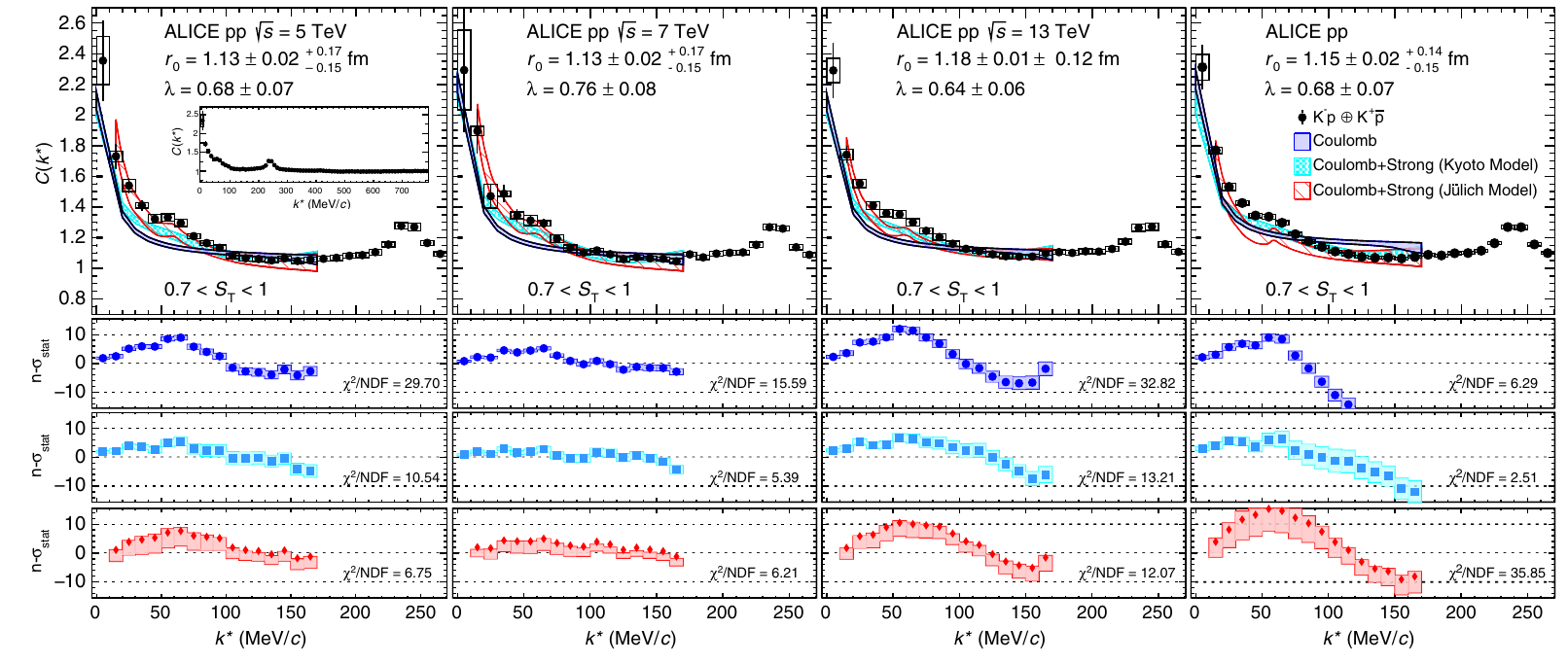}
  \caption{$(K^{-} p \oplus K^{+} \bar{p})$ correlation functions in proton-proton collisions at $\sqrt{s} = 5, 7, 13$~TeV (from left to right) by ALICE at LHC. The rightmost panel shows the result of the combined analysis of all collision energies. Adapted from Ref.~\cite{ALICE:2019gcn}.}
  \label{fig:ALICE}
\end{figure}

Theoretical calculation of the correlation function of a pair of hadrons without channel couplings can be performed using the Koonin-Pratt formula~\cite{Koonin:1977fh,Pratt:1986cc}:
\begin{align}
   C(\bm{q}) &= \int d\bm{r}\;
   S(\bm{r})|\Psi^{(-)}(\bm{q};\bm{r})|^{2} ,
   \label{eq:KPformula}
\end{align}
where $S(\bm{r})$ is the source function characterizing the shape and size of the hadron emitting source.  The scattering wave function $\Psi^{(-)}(\bm{q};\bm{r})$ of the hadron pair with relative coordinate $\bm{r}$ is determined by the hadronic final state interactions. It follows from Eq.~\eqref{eq:KPformula} that the measurement of the correlation function can serve two complementary purposes. First,  using a hadron pair with a well specified interaction,  the measured correlation function permits to extract information about the emission source $S(\bm{r})$ in the collision systems~\cite{Goldhaber:1960sf,ALICE:2020ibs}.  On the other hand,  if the source function can be estimated through other considerations,  one can extract the information of the two-body interaction in $\Psi^{(-)}(\bm{q};\bm{r})$ from the correlation function measurement. Utilizing the latter strategy, the STAR collaboration at RHIC~\cite{STAR:2014dcy} and the ALICE collaboration at LHC~\cite{ALICE:2018ysd,ALICE:2020mfd} have studied several baryon-baryon interactions, which are difficult or impossible to measure in standard scattering experiments. 

Measurements of the $K^{-}p$ correlation functions in proton-proton collisions at $\sqrt{s} = 5, 7, 13$~TeV were reported in Ref.~\cite{ALICE:2019gcn}. In the analysis, the data of the $K^-p$ pair and its charge conjugation $K^+\bar{p}$ are combined (Fig.~\ref{fig:ALICE}). A remarkable fact is the non-monotonic behavior of the correlation function at the relative momentum $\sim 58$ MeV. Because this momentum corresponds to the threshold energy of the $\bar{K}^0n$ pair, this behavior can be interpreted as a threshold cusp effect. The threshold energy difference between $K^-p$ and $\bar{K}^0n$ is about 5 MeV and caused by isospin symmetry breaking.  A $K^{-}p$ scattering experiment at such low momentum is technically difficult.  The non-optimal resolution of the total cross section data at such low energy prohibited an experimental signal of the $\bar{K}^0n$ threshold cusp.  In contrast, the high precision of the correlation function data in Fig.~\ref{fig:ALICE} clearly show the $\bar{K}^0n$ cusp effect.  In addition, there are data points even below the $\bar{K}^0n$ threshold, corresponding to $K^{-}p$ scattering at extremely low energy.  The $K^{-}p$ correlation function can therefore provide an additional new constraint on the properties of the $\Lambda(1405)$.

\begin{figure}[tbp]
  \centering
  \includegraphics[width=7cm, bb=0 0 375 221]{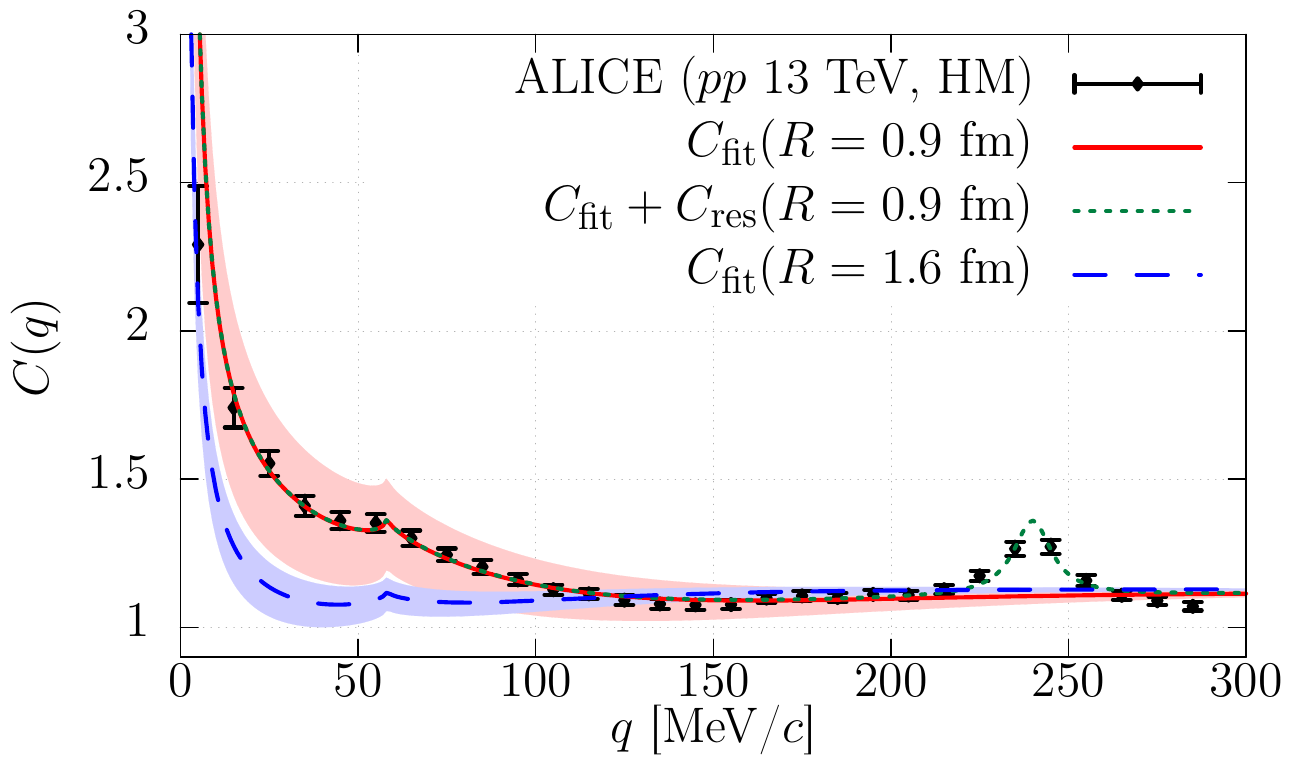}
  \caption{The $K^{-} p$ correlation function deduced from $pp$ collisions at $\sqrt{s} =13$~TeV (data) in comparison with theoretical calculation based on chiral $SU(3)$ dynamics with coupled channels (solid line).  Dotted line includes the contribution of the $\Lambda(1520)$ with $S=-1$ and $J^{P}=3/2^{-}$. Dashed line represents the prediction of the correlation function with larger source size. Adapted from Ref.~\cite{Kamiya:2019uiw}.}
  \label{fig:Correlation}
\end{figure}

An improved calculation of the $K^{-}p$ correlation function requires generalizing Eq.~\eqref{eq:KPformula} by taking into account the following items:
\begin{itemize}

\item Coupled-channel effects: the $K^{-}p$ system couples to the $\pi\Sigma$ and $\pi\Lambda$ channels at lower energies,  so a generalization of the Koonin-Pratt formula for the coupled-channel case is needed, as discussed in Refs.~\cite{Lednicky:1998r,Haidenbauer:2018jvl}.

\item Isospin symmetry breaking: to describe the $\bar{K}^{0}n$ cusp, the isospin symmetric formulation is not suitable.  Physical hadron masses with inclusion of isospin symmetry breaking must be used as input. 

\item Coulomb interaction: in the diagonal $K^{-}p$, $\pi^{+}\Sigma^{-}$, and $\pi^{-}\Sigma^{+}$ potentials, the Coulomb force is active in addition to the strong interaction, and the wave function components must be calculated for these channels with the appropriate asymptotic boundary conditions for Coulomb wave functions.

\end{itemize}
A theoretical framework including all these effects has been completed in Ref.~\cite{Kamiya:2019uiw}, and applied to the $K^{-}p$ correlation function. In this work, the meson-baryon coupled-channel potential based on chiral $SU(3)$ dynamics~\cite{Miyahara:2018onh} was used for the strong interaction to calculate the wave function. The source function $S(\bm{r})$ is assumed to be of spherical gaussian shape with a size parameter $R$. In Ref.~\cite{ALICE:2019gcn}, the source size is estimated as $R\sim 1$ fm from the $K^{+}p$ correlation function where the interaction is relatively well understood. By fixing $R=0.9$ fm and varying the weight of the $\pi\Sigma$ source,  $\omega_{\pi\Sigma}$, it is shown that the ALICE data can be well described (Fig.~\ref{fig:Correlation}). Once the theoretical framework is set, it is also possible to study the source size dependence of the correlation functions by changing the value of $R$. The prediction for a larger source is shown by the dashed line in Fig.~\ref{fig:Correlation}.  The correlation function in the small-momentum region is seen to be suppressed when the source size is increased. This tendency is consistent with the recent experimental result of Pb-Pb collisions reported by the ALICE collaboration in Ref.~\cite{ALICE:2021szj}.

\section{\textit{$\bar{K}N$ interaction and few-body kaonic nuclei}}

Let us now turn to kaon-nuclear few-body physics.  Several methods have been established to solve the few-body Schr\"odinger equation rigorously~\cite{Suzuki:1998bn,Kamada:2001tv,Hiyama:2003cu}. These methods permit to calculate few-body systems accurately for given interaction potentials. Realistic nuclear forces have been constructed by phenomenological potentials~\cite{Wiringa:1994wb,Machleidt:2000ge} and more recently based on chiral effective field theory~\cite{Epelbaum:2008ga,Machleidt:2011zz}. These potentials achieve an accuracy of $\chi^{2}/{\rm d.o.f}\sim 1$.  At the same time recent developments in lattice QCD~\cite{Ishii:2006ec,Aoki:2009ji} are promising in providing yet another approach to the $NN$ interaction by generating these potentials from the underlying QCD.  

Studies of few-body kaonic nuclei rely on realistic $\bar{K}N$ interactions in combination with advanced few-body techniques.  Such calculations have been performed using variational methods~\cite{Yamazaki:2007cs,Dote:2008in,Dote:2008hw,Barnea:2012qa,Ohnishi:2017uni} or solving Faddeev equations~\cite{Shevchenko:2006xy,Shevchenko:2007ke,Ikeda:2007nz,Ikeda:2010tk,Shevchenko:2014uva,Revai:2014twa}.  The most advanced computations~\cite{Ohnishi:2017uni,Shevchenko:2014uva,Revai:2014twa} use $\bar{K}N$ interactions constrained by experimental data including the SIDDHARTA measurement at the level of $\chi^{2}/{\rm d.o.f}\sim 1$.

\begin{figure}[tbp]
  \centering
  \includegraphics[width=7cm,bb=0 0 765 519]{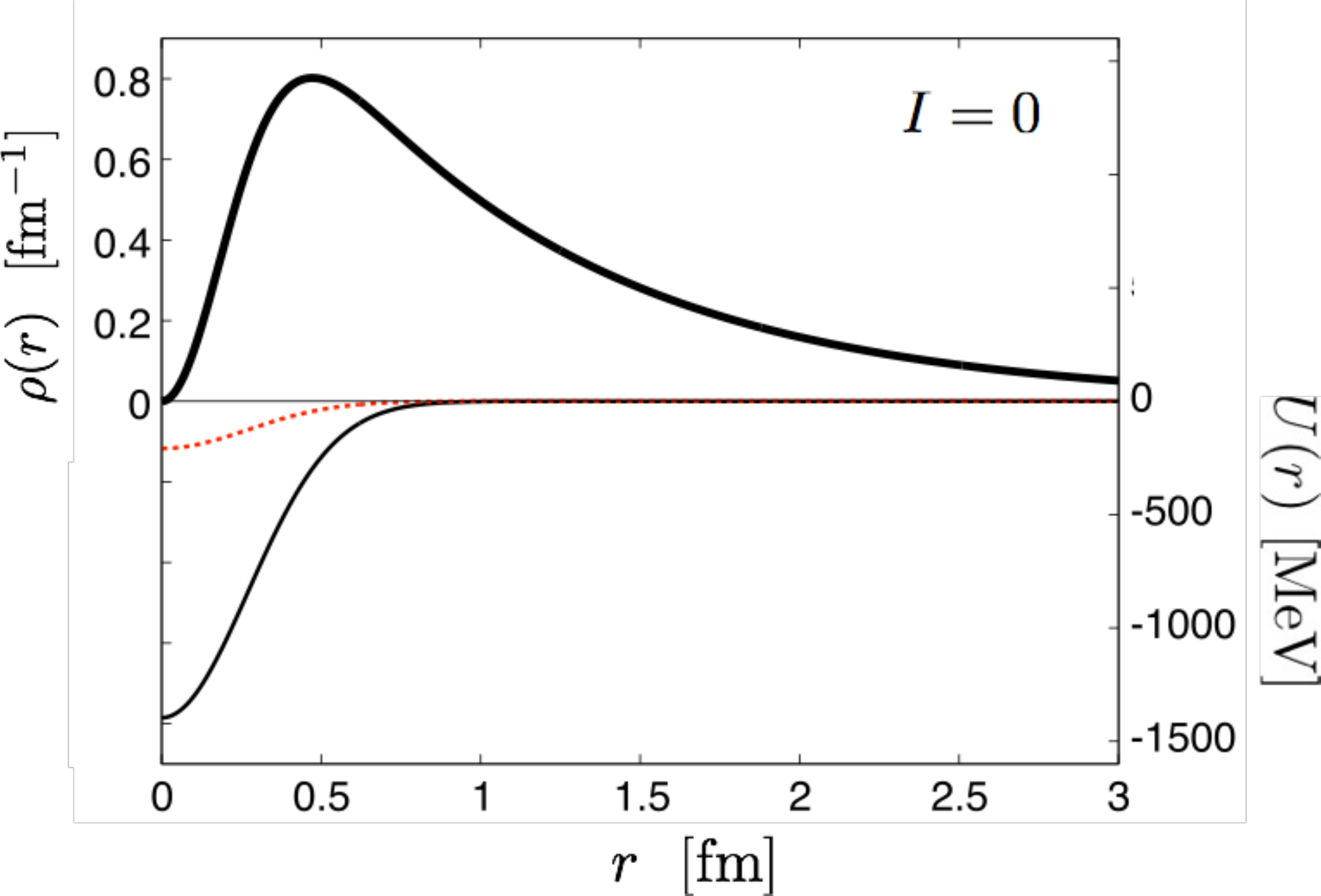}
  \caption{The $\bar{K}N$ density distribution $\rho(r)= r^2|\Psi(r)|^2$ (thick solid line) at the $\Lambda(1405)$ pole energy,  $(1424 - 26i)$ MeV. Also shown are the real part (thin solid line) and the imaginary part (dotted line) of the Kyoto $\bar{K}N$ potential $(I=0)$ at that energy.  Adapted from Ref.~\cite{Miyahara:2015bya}.}
  \label{fig:L1405density}
\end{figure}

Such potentials (called Kyoto $\bar{K}N$ potentials) have been presented in Refs.~\cite{Miyahara:2015bya,Miyahara:2018onh}, following the prescription developed in Ref.~\cite{Hyodo:2007jq} to construct an equivalent local potential that produces the scattering amplitude resulting from NLO chiral $SU(3)$ dynamics. Based on this scattering amplitude~\cite{Ikeda:2011pi,Ikeda:2012au},  an effective single-channel $\bar{K}N$ potential has been constructed~\cite{Miyahara:2015bya}, in which the coupled-channel decays of $\bar{K}N$ to $\pi\Sigma$ and $\pi\Lambda$ are incorporated by the imaginary part of the $\bar{K}N$ potential. The energy dependence of the potential reflects the derivative coupling of the NG bosons in chiral perturbation theory and the elimination of the $\pi$-hyperon channels.  Alternatively,  coupled-channel $\bar{K}N$-$\pi\Sigma$-$\pi\Lambda$ potentials have also been constructed~\cite{Miyahara:2018onh}.  In this case,  all coupling strengths are real and the energy dependence is milder than that of the effective single-channel $\bar{K}N$ potential,  as the $\pi Y$ channels are explicitly active in this framework.  As far as the $\bar{K}N$ component is concerned,  both the single-channel and coupled-channel potentials give equivalent results,  while the coupled-channel potential is suitable for the computation of quantities for which the $\pi\Sigma$ and $\pi\Lambda$ channels become important,  such as the $K^{-}p$ correlation functions.  In addition, the coordinate space potential is useful to extract the spatial structure of the system by displaying the wave function.  In Fig.~\ref{fig:L1405density}, we plot the density distribution $\rho(r)=r^{2}|\Psi(r)|^{2}$ at the $\Lambda(1405)$ pole together with the real and imaginary parts of the $\bar{K}N$ potential in the $I=0$ channel at the same energy. It is seen that the two-body density spreads over a large $\bar{K}N$ distance,  even beyond the range of the potential. This indicates again the $\bar{K}N$ molecular nature of the $\Lambda(1405)$. 

The simplest prototye kaonic nucleus is the three-body $\bar{K}NN$ system. Before proceeding to the actual calculations,  let us first classify the possible configurations.  Both $\bar{K}$ and $N$ have isospin $I=1/2$, so we can construct the $\bar{K}NN$ system in three different isospin configurations~\cite{PL7.288}:
\begin{align}
   {\rm (i)}: |\bar{K}[NN]_{I=0}\rangle_{I=1/2}, \quad
   {\rm (ii)}: |\bar{K}[NN]_{I=1}\rangle_{I=1/2}, \quad
   {\rm (iii)}: |\bar{K}[NN]_{I=1}\rangle_{I=3/2} .
   \label{eq:KbarNNstates}
\end{align}
Assuming that all the pairs are combined in $s$ waves in the ground state,  the $[NN]_{I=0}$ ($[NN]_{I=1}$) pair is  in a spin $J=1$ ($J=0$) state. Then the states (i), (ii), and (iii) have different quantum numbers: $I(J^{P})=1/2(1^{-})$, $1/2(0^{-})$, and $3/2(0^{-})$, respectively.\footnote{The state (ii) is sometimes called $K^{-}pp$, but this assignment is not appropriate because the $K^{-}pp$ configuration mixes with $\bar{K}^{0}pn$. The physical state is the superposition of these two configurations.  Furthermore (ii) is an isospin doublet. The isospin partner of the $K^{-}pp$\,-\,$\bar{K}^{0}pn$ state is a superposition of $K^{-}pn$ and $\bar{K}^{0}nn$.} Let us now consider the lowest energy state among (i)-(ii). The $[NN]_{I=0}$ ($[NN]_{I=1}$) configuration corresponds to the $^{3}S_{1}$ ($^{1}S_{0}$) channel, so only (i) is bound in the absence of $\bar{K}$. However, by performing the isospin recombination, one can extract the fractions of the $\bar{K}N$ pairs in different isospin states:
\begin{align}
   {\rm (i)}: |\bar{K}[NN]_{I=0}\rangle_{I=1/2}
   &= 
   -\frac{1}{2}
   |[\bar{K}N]_{I=0}N\rangle_{I=1/2}
   +\frac{\sqrt{3}}{2}
   |[\bar{K}N]_{I=1}N\rangle_{I=1/2} ,\\ 
   {\rm (ii)}: |\bar{K}[NN]_{I=1}\rangle_{I=1/2}
   &= 
   \frac{\sqrt{3}}{2}
   |[\bar{K}N]_{I=0}N\rangle_{I=1/2}
   +\frac{1}{2}
   |[\bar{K}N]_{I=1}N\rangle_{I=1/2} , \\
   {\rm (iii)}: |\bar{K}[NN]_{I=1}\rangle_{I=3/2}
   &= |[\bar{K}N]_{I=1}N\rangle_{I=3/2} .
\end{align}
This means that the fraction of the $\bar{K}N$ pair in the $I=0$ channel is $1/4$, $3/4$, and $0$ for (i), (ii), and (iii), respectively. Now we recall that the $\bar{K}N$ pair in $I=0$ has a quasi-bound state $\Lambda(1405)$ below the threshold, while the $I=1$ component is unbound, though attractive (see Table~\ref{tbl:interactions}). Thus, from the viewpoint of the $\bar{K}N$ interaction, the configuration (ii) can earn the largest attraction where the fraction of the $I=0$ $\bar{K}N$ pair is maximal. In this way, the $NN$ correlation favors the state (i), while the $\bar{K}N$ correlation favors the state (ii) as the ground state. The numerical calculation shows that there exists a quasi-bound state in channel (ii), while (i) is unbound with respect to the $K^{-}d$ threshold~\cite{Barnea:2012qa}. Thus, in the two-baryon systems, the $\bar{K}N$ correlation is so strong that the ground state spin is changed when a $\bar{K}$ is added to the nucleons.

Now we present numerical results of a comprehensive study of few-body kaonic nuclei in Ref.~\cite{Ohnishi:2017uni}, where the ground states of up to the seven-body system ($\bar{K}NNNNNN$) are computed using the stochastic variational method with a correlated gaussian basis~\cite{Suzuki:1998bn}. The AV4' potential~\cite{Wiringa:2002ja} is adopted for the nuclear force  which reproduces the few-body nuclei in the relevant mass number region $A=2$-6 reasonably well. The $\bar{K}N$ potential is the single-channel Kyoto potential from Ref.~\cite{Miyahara:2015bya} where the effects of the $\pi\Sigma$ and $\pi\Lambda$ channels are renormalized into the imaginary part of the potential. 

The results of the few-body calculations~\cite{Ohnishi:2017uni} are summarized in Table~\ref{tbl:binding}. Note that the absorption processes of the antikaon by multi-nucleons,  such as $\bar{K}NN\to YN$,  are not included (for estimates of such effects,  see Refs.~\cite{Dote:2008hw,Sekihara:2009yk,Sekihara:2012wj,Bayar:2013lxa}). Up to $A=4$ systems, $\bar{K}NNNN$, we find a quasi-bound state in each mass number below the lowest threshold. With $A=5$, no state is found below the threshold of $(\bar{K}NNNN)+N$. In the $A=6$ system, both the $J^{P}=0^{-}$ and $1^{-}$ system have a quasibound state at almost degenerated energy. The ground state of the $\bar{K}NN$ system is confirmed to be $J^{P}=0^{-}$ (the state (ii) in Eq.~\eqref{eq:KbarNNstates}), and the other states are found to be unbound. The uncertainties of the binding energy $B$ and the mesonic decay width $\Gamma_{\pi YN}$ come primarily from the treatment of the energy dependence of the $\bar{K}N$ system in the few-body kaonic nuclei. 

\begin{table}[btp]
\caption{Isospin $I$, spin-parity $J^{P}$, binding energy $B$, and the mesonic decay width $\Gamma_{\pi YN}$ of the ground states of the few-body kaonic nuclei~\cite{Ohnishi:2017uni}. Uncertainties stem mainly from the treatment of the energy dependence of the $\bar{K}N$ interaction and also from the difference in the isospin multiplet. The $0^{-}$ and $1^{-}$ states of $\bar{K}NNNNNN$ are almost degenerate.}
\begin{center}
\begin{tabular}{|l|c|c|c|c|}
\hline
 & $\bar{K}NN$ & $\bar{K}NNN$ & $\bar{K}NNNN$ & $\bar{K}NNNNNN$ \\ \hline
  $I(J^{P})$
  & $1/2(0^{-})$
  & $0(1/2^{-})$
  & $1/2(0^{-})$
  & $1/2(0^{-},1^{-})$ \\
  $B$ [MeV] 
  & 25-28
  & 45-50 
  & 68-76
  & 70-81 \\
  $\Gamma_{\pi YN}$ [MeV]
  & 31-59
  & 26-70 
  & 28-75
  & 24-76 \\
\hline
\end{tabular}
\end{center}
\label{tbl:binding}
\end{table}%

As mentioned above, in the three-body system, the $\bar{K}N$ correlation overcomes the $NN$ correlation and the spin-parity of the ground state is changed by the injection of the antikaon.  Consider next the interplay between the $\bar{K}N$ and $NN$ correlations in the four-nucleon systems, the $\bar{K}NNNN$ state. The wavefunction of the $\bar{K}NNNN$ system with $J^{P}=0^{-}$, $I=1/2$, $I_{3}=+1/2$ can be written:
\begin{align}
   |\bar{K}NNNN\rangle
   &= 
   C_{1}|K^{-}pppn\rangle
   +C_{2}|\bar{K}^{0}ppnn\rangle ,
   \label{eq:KNNNNwf}
\end{align}
where $C_{i}$ is the weight of each component. The $\bar{K}N$ correlation tends to increase the fraction of the $I=0$ pair ($K^{-}p$ or $\bar{K}^{0}n$), and therefore the first term in Eq.~\eqref{eq:KNNNNwf} is favored. The $NN$ correlation, on the other hand, favors the second component, because $ppnn$ can form an $\alpha$ particle cluster which has a large binding energy.  In this way, the $C_{i}$ weight should be determined by the competition between the $\bar{K}N$ and $NN$ correlations which act in opposite directions. The results of the dynamical few-body calculation shows that $|C_{1}|^{2}=0.08$ and $|C_{2}|^{2}=0.92$: the $NN$ correlation (in particular the $\alpha$ particle correlation) wins over $\bar{K}N$ correlation in the $\bar{K}NNNN$ system. 

\begin{figure}[tbp]
  \centering
  \includegraphics[height=50mm,angle=-00,bb=0 0 1920 1080]{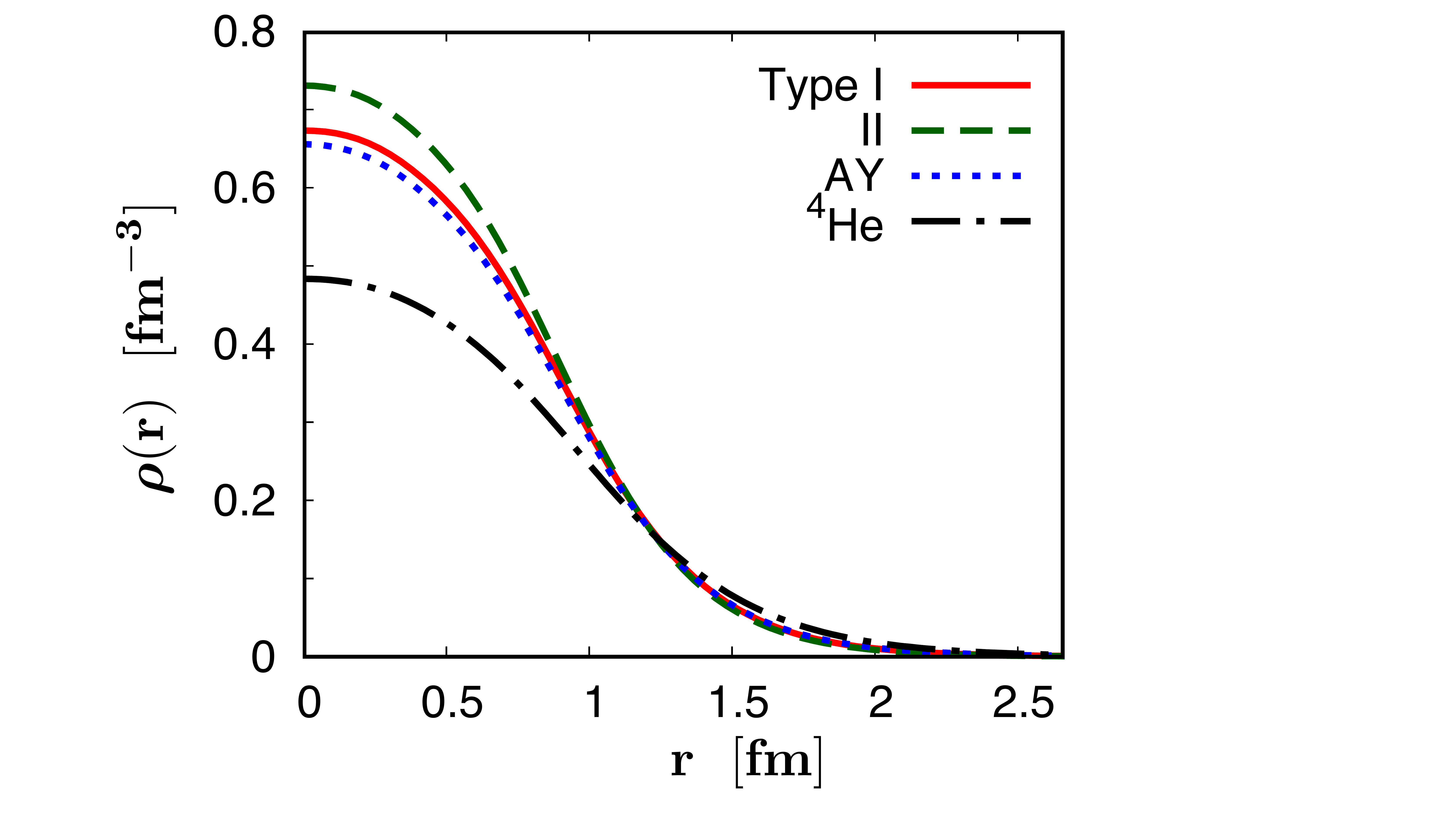}
  \caption{Nucleon density distribution in the $\bar{K}NNNN$ system measured from the center-of-mass of the nucleons by the Kyoto $\bar{K}N$ potential (solid and dashed lines) and by the Akaishi-Yamazaki potential (dotted line). For comparison, the density distribution of $^{4}$He is shown by the dash-dotted line. 
  Adapted from Ref.~\cite{Ohnishi:2017uni}.}
  \label{fig:KNNNN}
\end{figure}

From the relatively large binding energies of the kaonic nuclei in Table~\ref{tbl:binding}, one might expect that the density in the center of the system is also increased.  In fact early studies~\cite{Akaishi:2002bg} considered kaonic nuclei  to be candidates of possible high-density matter.  Here we comment on the density of the system from the viewpoint of the few-body calculation.  In Fig.~\ref{fig:KNNNN},  the density distributions of the nucleons in the $\bar{K}NNNN$ system measured from the center-of-mass of the nucleons are shown by the solid and dashed lines (the difference arises from the treatment of the energy dependence of the $\bar{K}N$ interaction). Compared with the density of $^{4}$He without the antikaon (dash-dotted line),  the central density increases slightly,  due to the attraction of the $\bar{K}N$ interaction.  For comparison, the $\bar{K}NNNN$ system is also calculated with the Akaishi-Yamazaki potential~\cite{Yamazaki:2007cs}. The binding energy of the $\bar{K}NNNN$ system is obtained as $B\sim 85$ MeV, larger than $B=68-76$ MeV with the Kyoto potential which is less attractive.  In Fig.~\ref{fig:KNNNN},  however,  the central density (dotted line) is smaller than the result of the Kyoto potential.  Thus, the larger binding energy does not necessarily imply a higher central density of the system. This is because the binding energy mainly reflects the tail of the wave function at large $r$.  In addition,  one should be cautious about the notion of the ``density'' in the few-body systems,  where in any case only a few nucleons are present.

Obviously,  remaining uncertainties concern the widths of $\bar{K}$-nuclear clusters with three and more nucleons.
The widths listed in Table~\ref{tbl:binding} are lower limits.  They do not include multinucleon absorption processes such as $\bar{K}NNN\rightarrow \Lambda NN$ and $\Sigma NN$ which become more likely with increasing number of nucleons, as the momentum mismatch characteristic of antikaon absorption on a two-nucleon pair can then be shared between those additional nucleons. 

Faddeev calculations of $\bar{K}NN$ systems have been performed~\cite{Shevchenko:2014uva,Revai:2014twa} using as input several $\bar{K}N$ potentials,  both phenomenological and chiral.  In search for $K^-pp$ bound clusters, binding energies and widths are found which are slightly larger than those listed in Table~\ref{tbl:binding}. A study comparing two-pole and and one-pole effective potential subject to SIDDHARTA constraints~\cite{Shevchenko:2012np} pointed out almost equivalent results with pole positions of the $\Lambda(1405)$ at  $1426-i 48$ MeV in the one-pole model,  and $1414-i58$ MeV and $1386-i104$ MeV in the two-pole model. 

In search for a $\bar{K}NN$ quasibound cluster, the J-PARC E15 collaboration measured the  $\Lambda p$ invariant mass spectrum in the reaction $^3{\rm He}(K^-,\Lambda p)n$ reaction~\cite{E15a:2019, E15b:2020}.  A binding energy $B = 42\pm 3 (stat.)^{+3}_{-4} (sys.)$ MeV was reported for the proposed $``K^-pp"$ cluster,  accompanied by a very large width, $\Gamma = 100\pm 7 (stat.)^{+19}_{-9}(sys.)$ MeV,  more than twice the binding energy.  This large width includes primary direct two- and three-nucleon absorption mechanisms,  $K^- (pp)n\rightarrow (Yp)n$ and $K^- (ppn) \rightarrow YNN$.  It raises,  however,  the issue whether the extremely short lifetime of such an object still justifies the notion of a bound system.

\section{\textit{Kaonic atoms}}

Kaonic atoms are systems of a negatively charged $K^{-}$ with positively charged ordinary nuclei,  bound by the Coulomb interaction. Although the $K^{-}$ interacts with nucleons also through the strong interaction, most of the energy levels of kaonic atoms follow the pattern of a hydrogen-like atom,  replacing the reduced mass of the system correspondingly. This is because the typical length scale of kaonic atoms is much larger than the range of the strong interaction.  Consider the simplest example of kaonic hydrogen, the $K^{-}p$ system. The Bohr radius of  kaonic hydrogen is $r_{B}\sim 84$ fm, which is much larger than the $K^{-}p$ interaction range of $\lesssim 1$ fm.  Nonetheless, the wave function of the ground state (the $1s$ level) of kaonic hydrogen has non-zero probability of the $K^{-}p$ pair to overlap with the strong-interaction range at short distance.  This induces  an energy shift,  $\Delta E$,  for the 1s level expected from the electromagnetic interaction alone,  and also a finite width $\Gamma$  of the energy level, reflecting the transitions $K^{-}p\to \pi\Sigma$ and $\pi\Lambda$.

In the study of pionic atoms, the Deser-Trueman formula~\cite{Deser:1954vq,Trueman:1961zz} was developed to relate the shift $\Delta E$ and the width $\Gamma$ to the two-body scattering length of the strong interaction. This formula has been further improved by including the isospin symmetry breaking effect by the effective Lagrangian approach as~\cite{Meissner:2004jr,Meissner:2006gx}
\begin{align}
   \Delta E - \frac{i\Gamma}{2} 
   &= -2\mu^2\alpha^3 a_{K^{-}p}
   \left[1-2\mu\alpha(\ln\alpha-1)a_{K^{-}p}\right]  ,
   +\dotsb,
   \label{eq:ImprovedDT}
\end{align}
with the reduced mass $\mu=m_{K^{-}}M_{p}/(m_{K^{-}}+M_{p})$, and the electromagnetic fine structure constant $\alpha\simeq 1/137$. The scattering length of the $K^{-}p$ system, $a_{K^{-}p}$,  is related to the $\bar{K}N$ scattering lengths in the isospin basis as 
\begin{align}
   a_{K^{-}p}
   &= \frac{1}{2}(a_{I=0}+a_{I=1})+\dotsb ,
   \label{eq:aKmp}
\end{align}
where the ellipsis stands for the isospin symmetry breaking effects. In Ref.~\cite{Baru:2009tx},   a further improvement was suggested by a resummation of logarithmically enhanced correction terms as
\begin{align}
   \Delta E-\frac{i\Gamma}{2}
   =-\frac{2\mu^{2}\alpha^{3}a_{K^{-}p}}{1+2\mu\alpha(\ln\alpha-1)a_{K^{-}p}}
   +\dotsb .
   \label{eq:resummedDT}
\end{align}

As already mentioned,  accurate kaonic hydrogen measurements were performed by the 
SIDDHARTA collaboration~\cite{SIDDHARTA:2011dsy,Bazzi:2012eq}. The 1s energy shift and width deduced from this experiment, $\Delta E =  -283\pm 36\pm 6$ eV and $\Gamma = 541\pm 89\pm 22$ eV,  translate through the Deser formula~\eqref{eq:ImprovedDT} into the complex $K^- p$ scattering length shown in Fig.~\ref{fig:amplitude}. 

Experiments measuring the 1s energy shift and width of kaonic deuterium are the next important step in order to further constrain the $\bar{K}N$ interactions,  as they provide additional sensitive access to the $I = 1$ component through the $K^-n$ interaction.  Such measurements are in preparation (SIDDHARTA-2 at DA$\Phi$NE~\cite{Curceanu:2019} and J-PARC E57~\cite{Zmeskal:2015}).  The theoretical description of kaonic deuterium is challenging because there is no principal separation of scales to start from.  The $K^- d$ Bohr radius of about 55 fm and the strong-interaction distance scale of about 1 fm have to be dealt with at the same level of high computational precision in order to achieve accurate results for the ground state energy shift and width.  Calculations of kaonic deuterium have been performed~\cite{Hoshino:2017mty} solving the three-body Schr\"odinger equation using a variational method.  The Kyoto $\bar{K}N$ potential~\cite{Miyahara:2015bya} has been used as input. The predicted energy shift and width of kaonic deuterium is
\begin{align}
   \Delta E-\frac{i\Gamma}{2} =(670-i508)\ {\rm eV}~~,
\label{eq:Kdshift}
\end{align}
with an uncertainty of about 10\% implied by the so far unconstrained $K^-n$ $(I = 1)$ interaction.  A kaonic deuterium measurement with 25\% accuracy is expected to improve the determination of this $I=1$ component significantly.  

A further observation concerns the applicability of the modified Deser-Trueman formulae,  Eqs.~\eqref{eq:ImprovedDT} and  \eqref{eq:resummedDT}. While these relations (especially the resummed version) work well for kaonic hydrogen, they are not reliable for kaonic deuterium.  A benchmark comparison can be made with the $K^-d$ scattering length,
\begin{align}
   a_{K^-d} = (-1.42 + i 1.60)\ {\rm fm}~~,
\end{align}
calculated in fixed-scatterer approximation using the full $K^-d$ Schr\"odinger wave function.  With this value of $a_{K^-d}$ the estimated $\Delta E$ resulting from \eqref{eq:resummedDT} deviates by more than 20\% from Eq. \eqref{eq:Kdshift}.  The deviation even exceeds 35\% for the improved Deser formula \eqref{eq:ImprovedDT}.  Hence the Deser formulae are not reliable for kaonic deuterium.  Faddeev calculations~\cite{Revai:2016muw} have arrived at similar conclusions.

Heavier kaonic atoms include nuclear mass numbers from A = 4 to A = 63,  systematically covering energy shifts and widths of kaonic 2p,  3d and 4f levels.  This systematics has been useful in establishing a phenomenological $K^-$-nuclear optical potential~\cite{Friedman:2007zza}.  More recent studies connect this optical potential to microscopic theoretical input for the $\bar{K}N$ interaction and address at the same time the role of nuclear $K^-$ absorption~\cite{Friedman:2016rfd}.  

The complex $K^-$- nuclear potential $U$ enters the Klein-Gordon equation describing the kaonic atom:
\begin{align}
\big[\vec{\nabla}^2 + \big(E-V_c(\vec{r})\big)^2-m_K^2-2\mu_K\,U(\vec{r})\big]\phi(\vec{r}) = 0~~,
\label{eq:KG}
\end{align}
where $\mu_K = m_K M_A/(m_K+M_A)$ is the $K^-$-nucleus reduced mass and $V_c$ is the Coulomb potential. In the limit of low proton and neutron densities, $\rho_p$ and $\rho_n$,  the leading order s-wave optical potential reduces to
\begin{align}
U(\vec{r}) = -{2\pi\over\mu_K}\left[{\cal F}_{K^-p}\,\rho_p(\vec{r}) + {\cal F}_{K^-n}\,\rho_n(\vec{r})\right]~~,
\label{eq:optpot}
\end{align}
with the $K^-N$ forward scattering amplitudes in the kaon-nucleus c.m. frame,  ${\cal F}_{K^-N} = \left(1 + {A-1\over A}{\mu_K\over M_N}\right)F_{K^-N}$ in terms of the threshold $K^-$nucleon c.m. amplitudes.  This leading-order single-nucleon potential is now commonly constructed using realistic theoretical input from chiral $SU(3)$ dynamics,  but it still needs to be extended by additional phenomenological pieces to account for correlations in the nucleus and for kaon absorption on two and more nucleons.

\begin{figure}[tbp]
  \centering
  \includegraphics[height=55mm,angle=-00,bb=0 0 1920 1080]{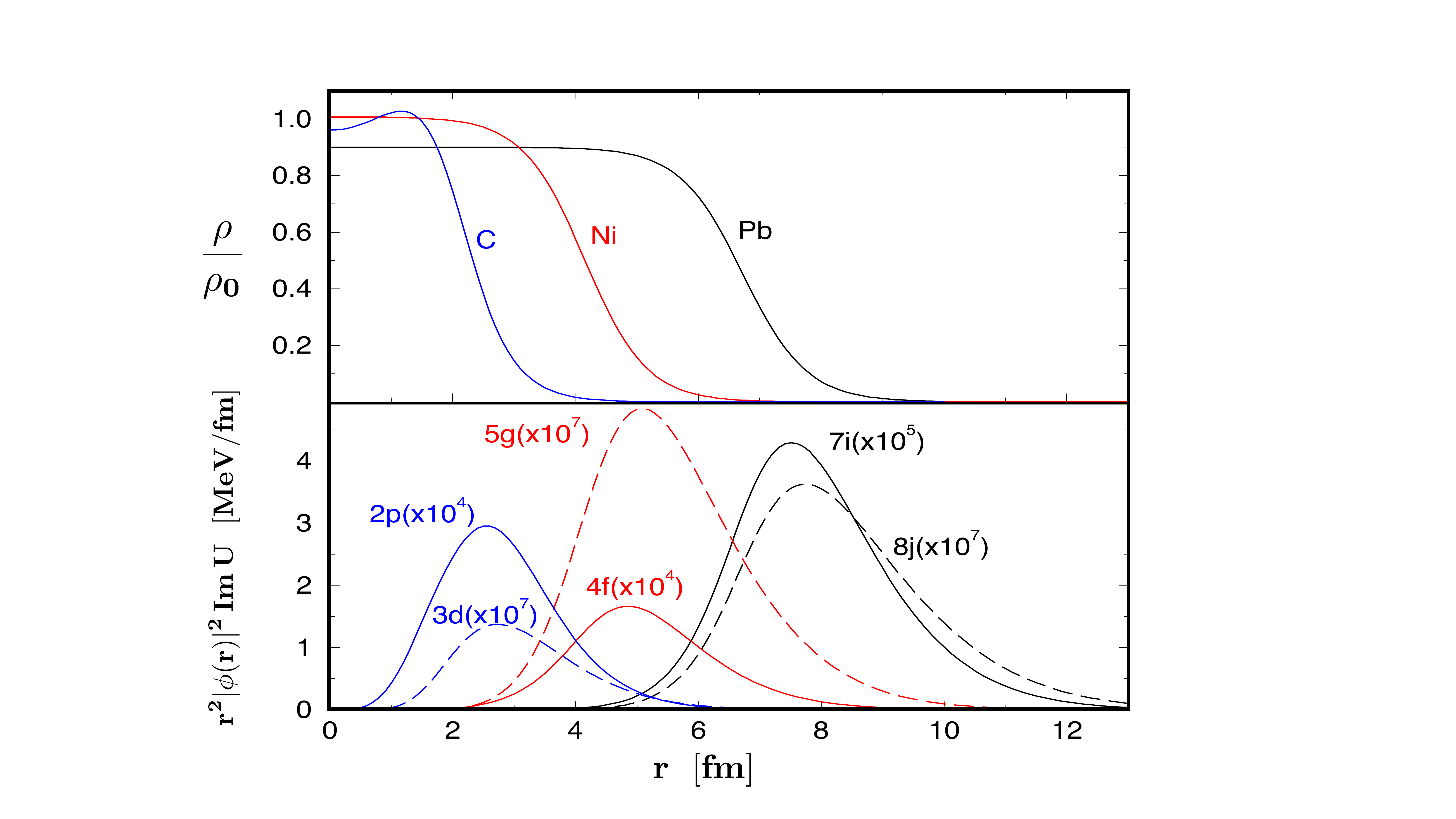}
  \caption{Absorptive interaction densities $r^2|\phi(r)|^2\,\textrm{Im}\,U(r)$ for several kaonic atoms for lower (solid curves) and upper (dashed curves) atomic levels in those selected examples.  Also shown are the corresponding density distributions (in units of $\rho_0 = 0.17$ fm$^{-3}$) of several atomic nuclei.  Adapted from Ref.~\cite{Friedman:2016rfd}.}
  \label{fig:absdensity}
\end{figure}

In Ref.~\cite{Friedman:2016rfd} a systematic investigation has been performed using an optical potential ansatz $U = U^{(1)} + U^{(2)}$,  where $U^{(1)}$ is based on Eq. \eqref{eq:optpot} using realistic chiral $SU(3)$ coupled-channel amplitudes as input,  but with additional corrections for subthreshold kinematics and Pauli principle effects~\cite{Waas1997b} taken into account.  A phenomenological second term $U^{(2)}$ is introduced to represent multinucleon processes,  in particular $K^-$ absorption:
 \begin{align}
U^{(2)}(\vec{r}) = -{2\pi\over\mu_K}B\,\rho(\vec{r})\left({\rho(\vec{r})\over\rho_0}\right)^\alpha~~,
\label{eq:U2}
\end{align}
with a complex strength $B$ and exponent $\alpha \geq 1$,  and the density $\rho = \rho_p + \rho_n$ scaled in units of the ground state density of nuclear matter, $\rho_0 = 0.17$ fm$^{-3}$.  A quantity of interest in this context is the fraction of single- and multi-nucleon absorption rates.  Single nucleon absorption refers to the $K^-N\rightarrow\pi\Lambda, \pi\Sigma$ reactions encoded in the imaginary parts of the amplitudes $F_{K^-N}$,  for which a fraction of $0.75\pm 0.05$ is quoted in Ref.~\cite{Friedman:2016rfd}.  The remaining 25\% are empirically assigned to multinucleon absorptive processes,  with no significant dependence on nuclear mass number.  Given this constraint,  an instructive outcome of a precision fit to a large kaonic atom data base is that,  out of a variety of $\bar{K}N$ potentials,  only two which are derived from chiral $SU(3)$ dynamics~\cite{Ikeda:2012au, Cieply:2012} actually qualify as being compatible with the data.

When combined with the single-nucleon potential $U^{(1)}$ constructed with chiral $SU(3)$ amplitudes~\cite{Ikeda:2012au},  a best fit for the strength of the phenomenological potential \eqref{eq:U2},  with fixed $\alpha = 1$, arrives at an absorption length, $B = (-0.9 +i\, 1.4)$ fm,  with an uncertainty of about 20\%.  The $K^-$ absorption processes are localized at the nuclear surface, as shown in Fig.~\ref{fig:absdensity}.  One can distinguish between absorption from a lower and an upper level of a given kaonic atom as indicated. The maximum overlaps of the absorption profiles with the nuclear density distributions are at relatively low densities, around $0.15 - 0.2\,\rho_0$ for the lower state and $0.1 - 0.15 \,\rho_0$ for the upper state~\cite{Friedman:2016rfd}.

\section{\textit{Many-body physics and kaons in baryonic matter}}

Kaon physics in the context of the nuclear many-body problem includes a broad range of topics,  from kaon- and antikaon scattering and production on complex nuclei to more exotic phenomena such as kaon condensation in dense baryonic matter and the possible role of strangeness in neutron stars. These themes have been addressed in several review articles~\cite{Ramos2001,Tolos:2020aln}.  In this section, after some short introductory digressions,  we shall primarily report on theoretical explorations in quest for possible quasibound antikaon-nuclear systems with more complex nuclei, following the previous assessment of few-body kaonic nuclei.

Historically,  the attractive nature of the $\bar{K}N$ interaction as it follows from the leading Weinberg-Tomozawa amplitude provided a strong motivation for investigating the phenomenon of kaon condensation in dense baryonic matter~\cite{Kaplan1986}.  The bulk of neutron star matter is traditionally viewed as being composed of neutrons plus a small fraction of protons and electrons subject to beta equilibrium.  At some critical density, a rapidly dropping in-medium $K^-$ mass might cross the continuously increasing electron chemical potential.  At that point it begins to be energetically favorable for the $K^-$ to replace the electrons,  and a kaon condensate is formed~\cite{Brown:1994}.  However, such a process would soften the equation-of-state of neutron star matter,  lowering its maximum mass to typically less than 1.5 solar masses.   Within the last decade,  new stringent conditions for the stiffness of the neutron star equation-of-state have been established by the observation of several heavy neutron stars with masses $M \simeq 2 \,M_\odot$ and even beyond.  As a consequence kaon condensation in neutron stars can be ruled out.  At the same time the discussion about the role of strangeness in neutron stars has turned to a new direction under the heading of the ``hyperon puzzle",  the fact that the present maximum mass constraints can also not easily be reconciled with possible admixtures of hyperons in the neutron star core~\cite{Djapo:2010,Logoteta:2019,Gerstung:2020}.

The starting point for studying interactions of antikaons in a nuclear medium with proton and neutron densities $\rho_p$ and $\rho_n$ (baryon density $\rho = \rho_p + \rho_n$) is the dispersion equation (energy $E$ and momentum $\vec{q}$ ) in homogeneous nuclear matter:
\begin{equation}
E^2 - \vec{q}^2 - m_K^2 - \Pi(E,\vec{q};\rho_p,\rho_n)=0~~,
\label{eq:disp}
\end{equation}
where $m_K$ is the $\bar{K}$ mass in vacuum and the self-energy $\Pi$ summarizes all interactions of the antikaon with nucleons in the medium.  A corresponding optical potential $U$ is defined as
\begin{equation}
U(E,\vec{q};\rho_p,\rho_n) ={1\over 2E} \Pi(E,\vec{q};\rho_p,\rho_n)~~.
\end{equation}
The energy and momentum dependent potential $U$ for antikaons is in general complex. Its imaginary part quantifies the absorption of the meson in the medium,  as already encountered in the case of kaonic atoms. At zero momentum, $\vec{q} = 0$,  the self-consistent solution of Eq.~\eqref{eq:disp} determines an effective in-medium mass of the meson:
\begin{equation}
 m_K^*(\rho) = \textrm{Re}\,E(\vec{q} = 0;\rho)~~.
\label{eq:effmass}
\end{equation}  
Correspondingly,  the in-medium meson decay width is given by
\begin{equation}
\Gamma(\rho) = -2\,\textrm{Im}\,E(\vec{q} = 0;\rho) \simeq -2\,\textrm{Im}\,U(m_K^*, \vec{q} = 0;\rho)~~,
\label{eq:width}
\end{equation} 
where the approximation on the r.h.s. holds if the width is small compared to the effective mass.

Consider now the self-energies of $K^-$ and $K^+$ in a nuclear medium.  In the low-density limit,
\begin{align}
\Pi^{\pm}= -4\pi{\sqrt{s}\over M_N}\left( F_{K^{\pm}p}\,\rho_p + F_{K^{\pm}n}\,\rho_n\right)~~,
\label{eq:Kselfenergy}
\end{align}
 in terms of the $K^{\pm}N$ forward scattering c.m. amplitudes.  At low energies close to threshold these amplitudes are governed  by chiral symmetry and the Weinberg-Tomozawa theorem as in Eq. \eqref{eq:VWT}:
\begin{align}
 F^{WT}_{K^-p}=- F^{WT}_{K^+p} = {E\over 4\pi\,f^2}~,~ ~~~~~
 F^{WT}_{K^-n}=- F^{WT}_{K^+n} = {E\over 8\pi\,f^2}~~,
\label{eq:WTterms}
\end{align}
implying a characteristic downward (upward) shift of the effective $K^-$ ($K^+$) masses by their attractive (repulsive) s-wave interactions, respectively,  in the nuclear medium.  Such a splitting is actually observable in the rapidity distributions of $K^-$ and $K^+$ produced in $1-2$ AGeV heavy-ion collisions~\cite{Song:2021}.

It is instructive to compare the s-wave optical potentials for pions and kaons from the point of view of the underlying chiral symmetry breaking pattern.  In the chiral limit of massless $u,  d$ and $s$ quarks,  all interactions of both pions and kaons as Nambu-Goldstone bosons would strictly vanish at $E = \vec{q} = 0$.  For the pion,  the leading s-wave potential at threshold,  $U = -{2\pi\over m_\pi}b_0\rho$,  is determined by the isospin-even $\pi N$ scattering length  $b_0$ for which a chiral low-energy theorem as well as pion-nucleon data give a very small value. Next-to-leading order nuclear correlation effects are therefore important.  For kaons and antikaons the driving s-wave terms~\eqref{eq:WTterms} are much stronger at threshold ($E = m_K$),  where the comparatively large kaon mass reflects the explicit chiral symmetry breaking by the mass of the strange quark.  For example and orientation,  the Weinberg-Tomozawa potential for a $K^-$ in symmetric nuclear matter becomes\begin{align}
 U^{WT}_{K^-}= - {3(1+m_K/M_N)\over 8\, f_K^2}\,\rho \simeq -60 \,{\rho\over\rho_0}\,\textrm{MeV}~~,
\label{eq:WTpot}
\end{align}
using the empirical kaon decay constant $f_K \simeq 110$ MeV and $\rho_0 = 0.17$ fm$^{-3}$.

\begin{figure}[tbp]
  \centering
  \includegraphics[height=40mm,angle=-00,bb=0 0 1920 1080]{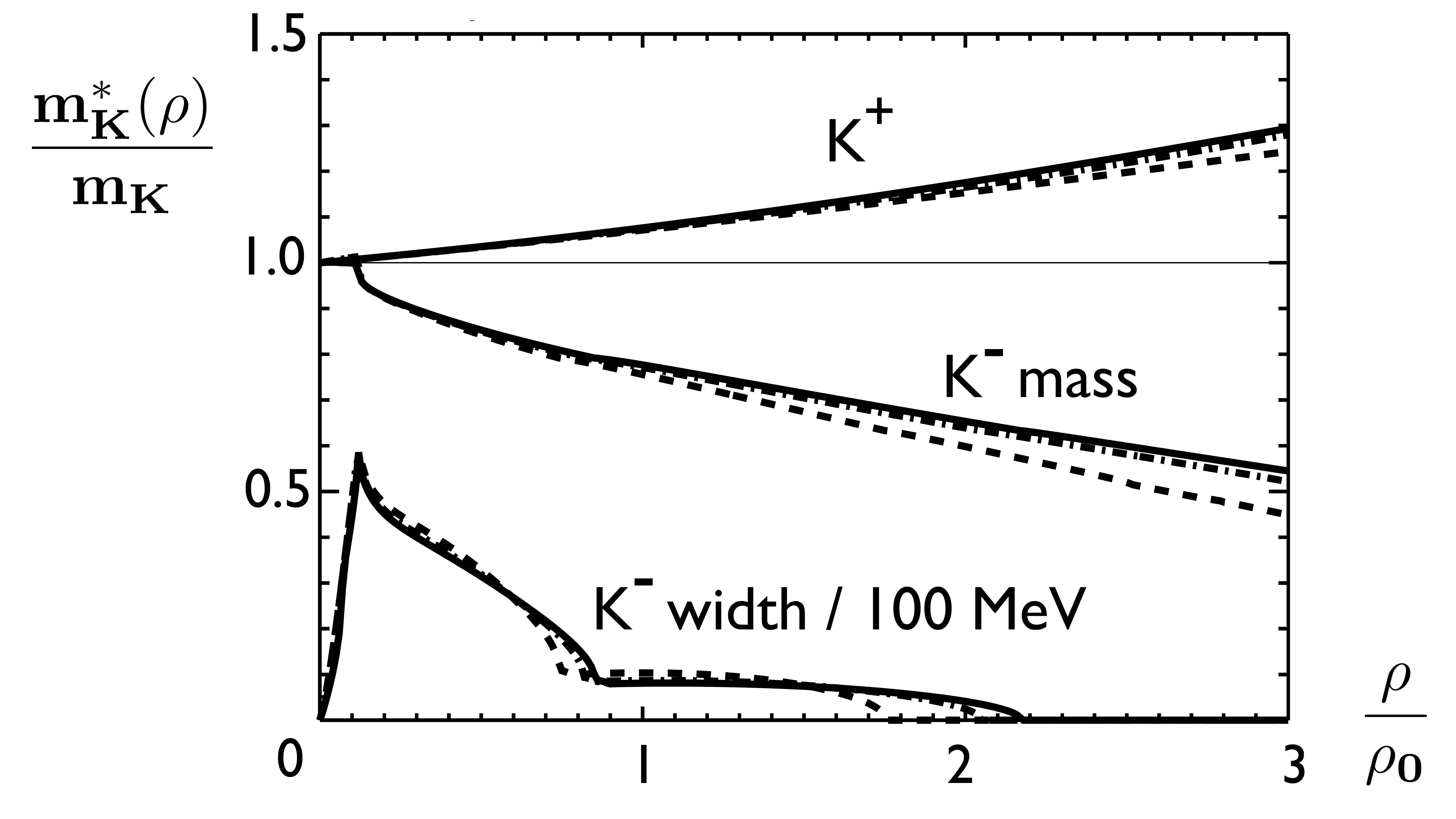}
  \caption{In-medium mass $m^*_K(\rho)$ and width of a $K^-$ in symmetric nuclear matter as function of baryon density $\rho$ in units of normal nuclear matter density,  $\rho_0 = 0.17$ fm$^{-3}$.  Calculations based on chiral $SU(3)$ dynamics with coupled channels and including effects of Pauli blocking and Fermi motion (dashed lines). Full lines: two-nucleon correlations additionally included.  Also shown is the in-medium $K^+$ mass for comparison. Adapted from Ref.~\cite{Waas1997b}.} 
  \label{fig:Kmass}
\end{figure}

Theoretical studies of kaons and antikaons in nuclear matter using amplitudes from chiral $SU(3)$ coupled-channel dynamics were reported in Refs.~\cite{Waas1996,Waas1997a, Waas1997b}.  Also included in these works were effects nonlinear in density,  i.e.  beyond the non-interacting nuclear Fermi gas,  in particular Pauli blocking of nucleons and short-range $NN$ correlations.  A typical result for the effective $K^-$ and $K^+$ masses is shown in Fig.~\ref{fig:Kmass}.  The primary Weinberg-Tomozawa mass splitting pattern between antikaon and kaon in matter is still prominent.  Pauli blocking effects are next in importance in shifting this pattern,  whereas short-range correlations turn out to have moderate influence and enter significantly at densities above twice the density of normal nuclear matter.  The large width shown in the lower part of Fig.~\ref{fig:Kmass} reflects the disappearance of the $K^-$ in the processes $\bar{K}N\rightarrow \pi Y$ with $Y = \Lambda, \Sigma$.  This width should be taken as a lower limit because multinucleon absorption mechanisms such as $\bar{K}NN\rightarrow YN,  ~\bar{K}NNN\rightarrow YNN$, ...  have not been considered.  At very low densities the formation of the $\Lambda(1405)$ and its in-medium subthreshold dynamics induces strong non-linear density dependence.  In Ref.~\cite{Lutz2002} this behaviour has been further elaborated by a self-consistent treatment of the $\Lambda(1405)$ and other hyperons in the nuclear medium.

\begin{figure}[tbp]
  \centering
  \includegraphics[height=40mm,angle=-00,bb=0 0 1920 1080]{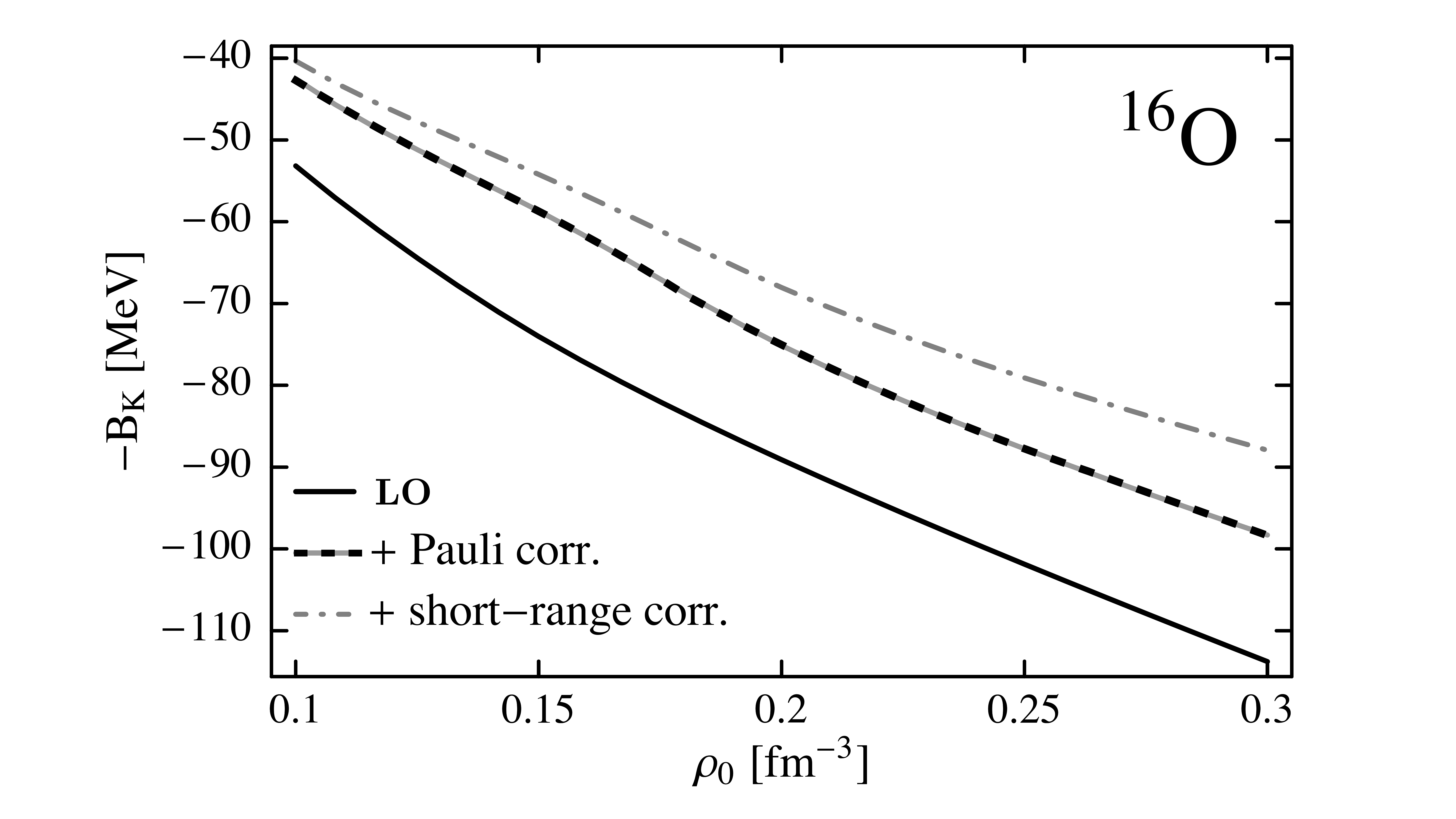}
  \includegraphics[height=38mm,angle=-00,bb=0 0 1920 1080]{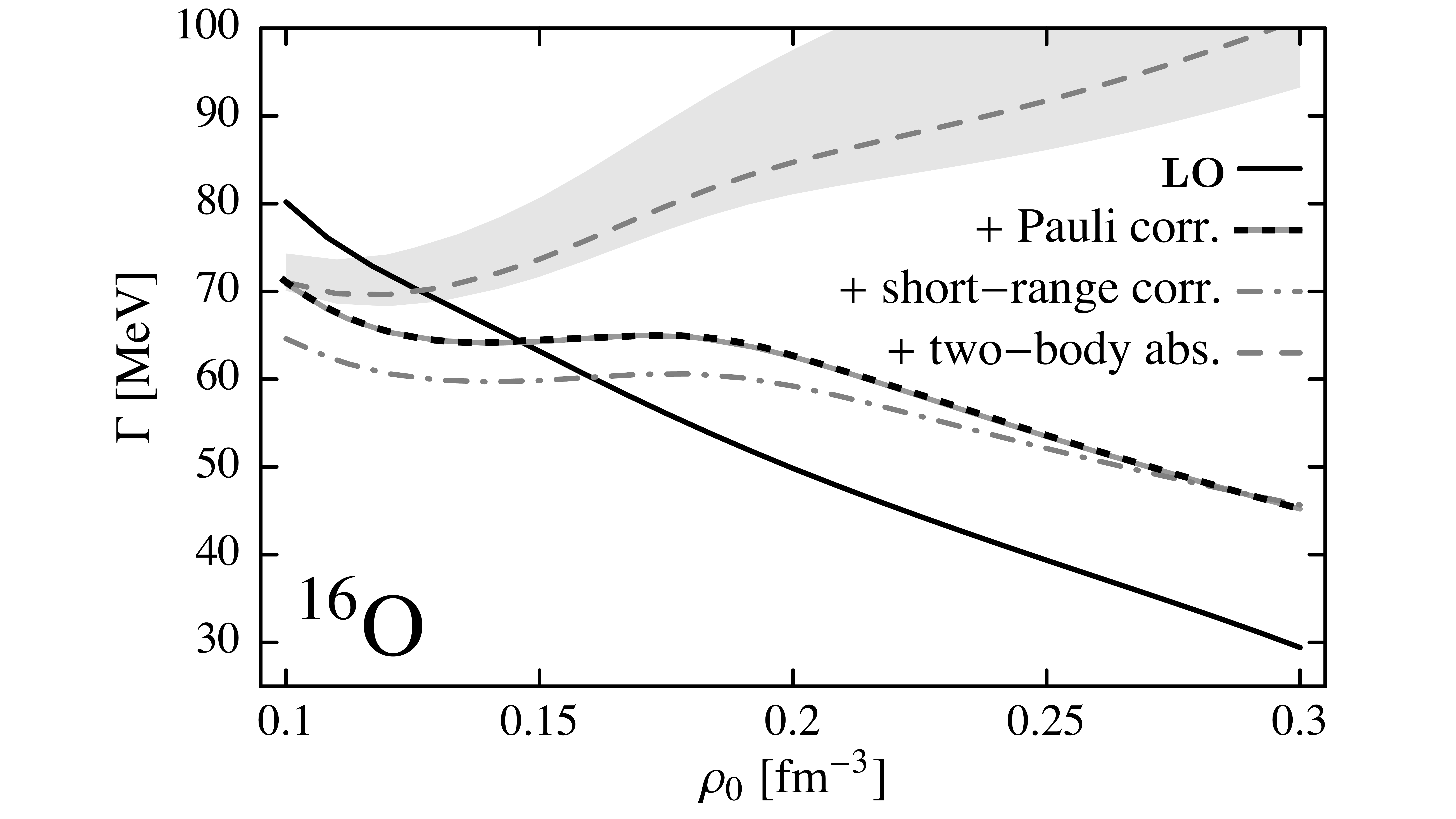}
  \caption{Binding energy $B_K$ and width $\Gamma$ of a $K^-$ bound in $^{16}{\rm O}$ as function of the central nuclear density $\rho_0$.  Solid curves (LO): leading order $s$- and $p$-wave interactions ($T\rho$ approximation). Short-dashed and dash-dotted curves: sequentially adding Pauli and short-range correlations.  Long-dashed curve with grey uncertainty band: estimated additional width from $\bar{K}NN$ two-body absorption.  Adapted from Ref.~\cite{Weise:2008aj}.}
  \label{fig:quasibound}
\end{figure}

The strong in-medium attraction experienced by the $\bar{K}$ has motivated several systematic studies of possible quasibound states in complex nuclei ~\cite{Weise:2008aj,Cieply:2011,Gazda:2012zz,Hrtankova:2017wrg}.  The investigation~\cite{Weise:2008aj} includes $s$- and $p$-wave $\bar{K}N$ interactions together with Pauli and short-range correlations,  plus an estimate of $\bar{K}NN$ two-nucleon absorption, in a calculation solving Eq.\,\eqref{eq:disp} with local nuclear densities, $\rho(\vec{r})$.  Binding energies $B_K$ and widths $\Gamma$ of the lowest antikaon quasibound states are computed for prepared pseudo-nuclei, varying the central density $\rho_0$ from its nominal value in order to explore the systematics of a possibly compressed $\bar{K}$-nuclear system.  An example is shown in Fig.~\ref{fig:quasibound} for the case of $^{16}{\rm O}+K^-$.  The strong binding in the leading ($T\rho$) approximation is significantly reduced by Pauli effects and,  to a lesser extent,  by additional short-range correlations.  The estimated two-body absorption width competes with the binding energy and begins to overwhelm $B_K$ as the central density increases,  even though this width is presumably still to be considered a lower limit since absorption processes on more than two nucleons add further contributions to the overall width.

\begin{figure}[tbp]
  \centering
  \includegraphics[height=60mm,angle=-00,bb=0 0 1920 1080]{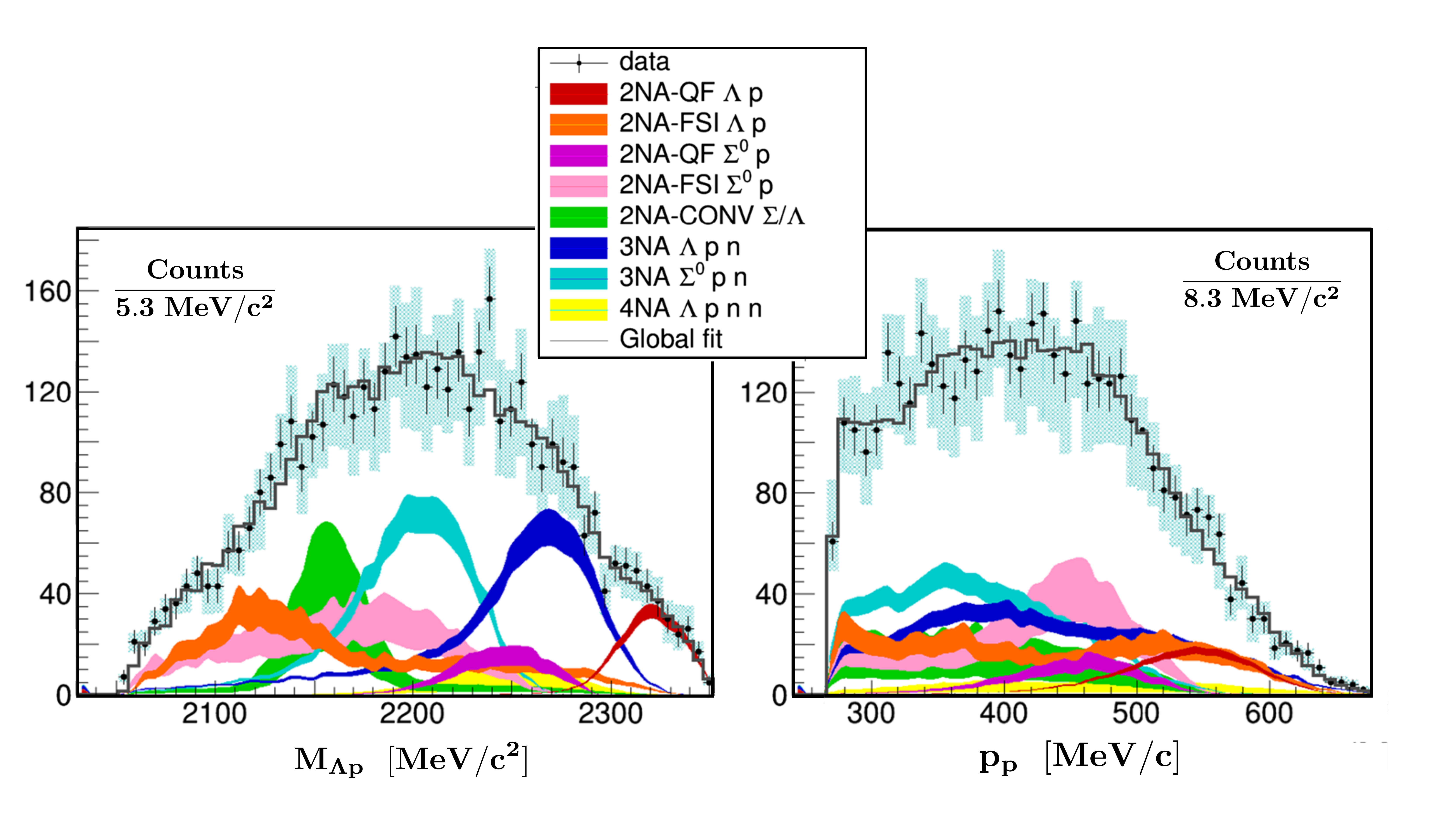}
  \caption{$\Lambda p$ invariant mass spectrum (left) and proton momentum distribution (right) following $K^-$ absorption on $^{12}C$ in the KLOE detector (LNF).  Color codes show contributions to total absorption from various processes involving two, three and four baryons.  Adapted from Ref.~\cite{DelGrande:2019}. }
  \label{fig:absorption}
\end{figure}

In this context,  a more detailed understanding of $\bar{K}$-nuclear systems clearly requires as much information as possible about the multi-nucleon absorption rates of antikaons through $\bar{K}NN \rightarrow YN,  ~\bar{K}NNN \rightarrow YNN, ...$ multi-nucleon processes, with hyperons $Y = \Lambda, \Sigma$ in the final state.  Experimental data have been reported in Refs.~\cite{VazquezDoce:2015szw,HADES:2018qkj, DelGrande:2019}.  Figure~\ref{fig:absorption} shows the $\Lambda p$ spectrum following low-momentum $K^-$ absorption on $^{12}\rm{C}$ layers in the KLOE detector at LNF.  These $\Lambda p$ spectra also incorporate conversions from $\Sigma^0 p$ final  states.  The hyperon-nucleon pairs are produced in various multi-nucleon absorption processes that are simulated to reconstruct the overall rate with an envelope extending over the region $2.1$-$2.3$ GeV,  covering as well the potential range of possible bound ``$K^-pp$" clusters.  Also shown in Fig.~\ref{fig:absorption} is the momentum spectrum of protons emerging from the absorption reactions.  One notes large proton recoil momenta characteristic of two-body absorption kinematics,  while three and more particles in the final state can share the momentum more evenly between them.  

\begin{figure}[tbp]
  \centering
  \includegraphics[height=60mm,angle=-00,bb=0 0 1920 1080]{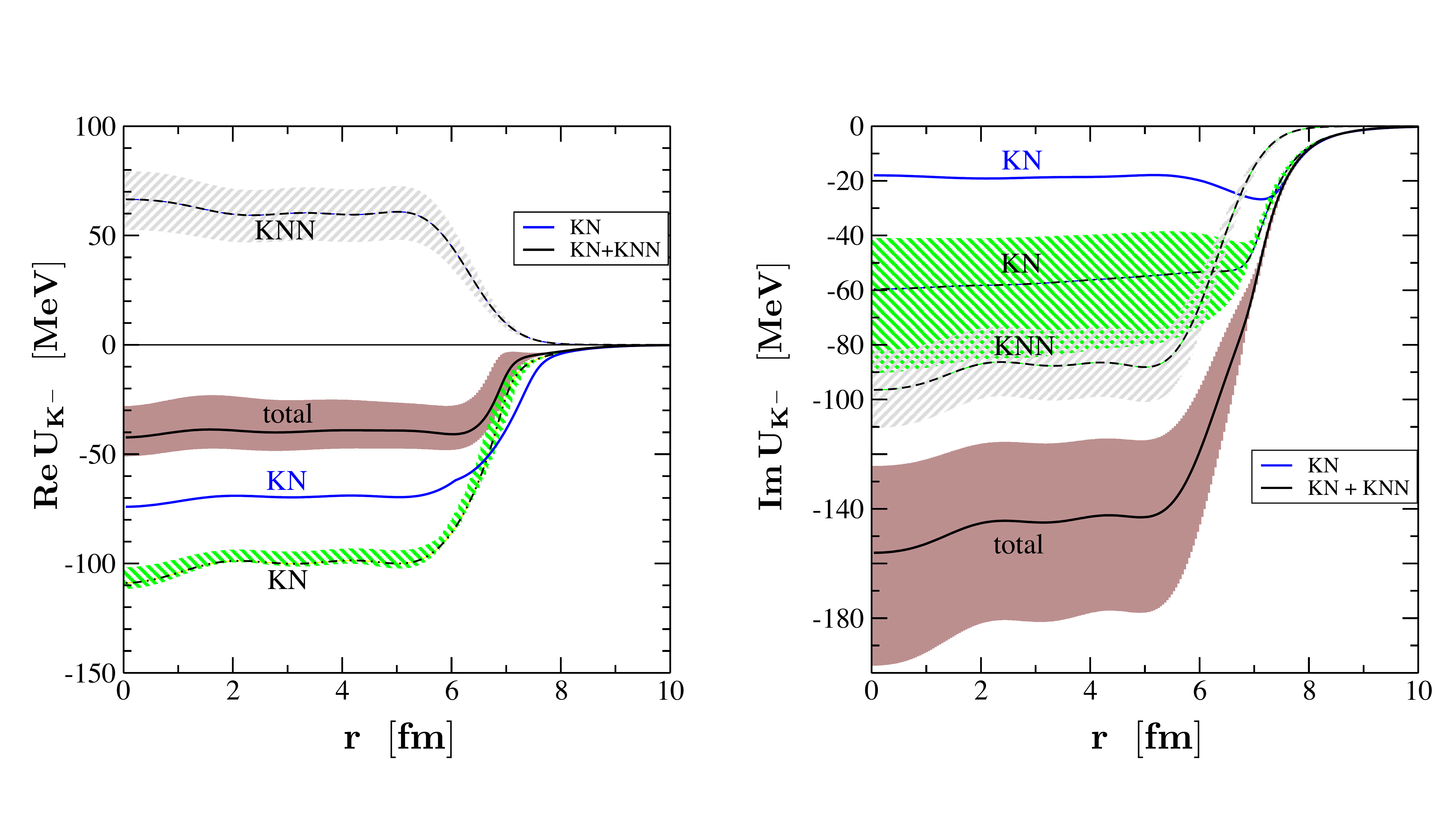}
  \caption{Real and imaginary parts of the $K^-$-nucleus potential for the $^{208}{\rm Pb}+K^-$ system.  The chiral $SU(3)$ dynamics based potential~\cite{Ikeda:2012au} has been used in combination with the phenomenological absorptive potential of Eq.\,\eqref{eq:U2}.  with $\alpha = 1$.  Different contributions from the $K^-N$ single-nucleon potential ($KN$,  including subthreshold kinematics) and the phenomenological multi-nucleon potential ($KNN$) are displayed,  together with the total potential including uncertainty estimates.   The chiral $K^-N$ potential (solid blue lines - in the absence of the multi-nucleon potential) is also shown for comparison.  Adapted from Ref.~\cite{Hrtankova:2017wrg}.} 
  \label{fig:Kbound}
\end{figure}

The competition between binding and absorption is obviously the key to the possible existence of quasibound $\bar{K}$-nuclear states. This issue has been further elaborated~\cite{Hrtankova:2017wrg} in an investigation that follows closely the constraints implied by the systematics of pionic atoms discussed in the previous section. The input is now again the single-nucleon chiral $SU(3)$-based coupled-channel potential plus the complex phenomenological multi-nucleon potential \eqref{eq:U2}.  The real and imaginary parts of the resulting potential that could, in principle,  support a quasibound $^{208}{\rm Pb}+K^-$ system is shown in Fig.~\ref{fig:Kbound}.  While the one-body potential derived from chiral $SU(3)$ dynamics would indeed be able to generate such a bound state,  the large imaginary part from multi-nucleon absorption, together with the accompanying repulsive real part, both work against binding.  The total real part of the potential adds up to a central strength,  Re~$U_K \simeq - 40$ MeV,  whereas the imaginary part turns out to be about four times as large,  Im~$U_K \simeq - 160 \pm 30$ MeV, though with considerable uncertainties. Nonetheless,  multinucleon absorption of antikaons is expected to reduce the lifetime of possible $\bar{K}$-nuclear quasibound states to a low level that will make such states difficult to observe.  The same kind of absorptive mechanisms would also work against kaon condensation in dense baryonic matter.

\section{\textit{Summary}}

The interest in strangeness nuclear physics is in large part motivated by the special role that the strange quark plays in low-energy QCD,  with its mass intermediate between light and heavy,  featuring a subtle combination of spontaneous and explicit chiral symmetry breaking.  This symmetry breaking pattern implies that the interaction of antikaons with nucleons close to threshold is quite strongly attractive.  As a consequence,  the question has been raised whether quasibound states of $\bar{K}$ with nucleons and nuclei can exist.  For the isospin $I =0$ channels of the $\bar{K}$-nucleon system,  this question found a convincing answer in the form of the $\Lambda(1405)$ as a $\bar{K} N$ quasibound state embedded in the $\pi\Sigma$ continuum with a nontrivial,  quasimolecular intrinsic structure.  Viewed in a simple constituent quark picture,  the $\Lambda(1405)$ can indeed be considered as historically the first example of an ``exotic" baryon with an underlying five-quark ($udu\bar{u}s$ and $udd\bar{d}s$) content.

The framework for the theoretical description of kaon- and antikaon-nuclear systems close to threshold is an effective field theory (EFT) based on spontaneously broken chiral $SU(3)_L\times SU(3)_R$ symmetry, as a realization of low-energy QCD with $N_f =3$ quark flavors.  Unlike the two-flavor case,  however, chiral perturbation theory is not an option to proceed because of the significant explicit chiral symmetry breaking by the strange quark mass and the appearance of the $\Lambda(1405)$ just below $\bar{K}N$ threshold. Instead,  a combination of this EFT with non-perturbative coupled-channel methods (chiral $SU(3)$ coupled-channel dynamics) proves to be successful in comparison with all known empirical $\bar{K}N$ data,  including accurate measurements of the atomic energy shift and width of kaonic hydrogen and the $K^-p$ correlation function from high-energy proton-proton collisions.  Important steps in the near future for further improvements include measurements of kaonic deuterium that are expected to provide more detailed information on the isospin $I = 1$ part of the $\bar{K}N$ interaction through the $K^-n$ channel.  

The systematics of strong interaction energy shifts and widths in kaonic atoms with heavier nuclei have always been an important source of constraints on the $K^-$-nuclear optical potential.  With the leading two-body $\bar{K}N$ interaction from chiral $SU(3)$ dynamics as a reliable input,  the focus on many-body effects such as $K^-$ absorption in kaonic atoms with heavier nuclei can now be better controlled.  

Given the strong attraction in the $\bar{K}N$ interaction that is manifest already in its leading-order (Weinberg-Tomozawa) term,  much scientific intensity has been devoted to theoretical and experimental investigations of possible bound $\bar{K}$-nuclear systems.  Several calculations have been addressed to the prototype $\bar{K}NN$ bound cluster and to light antikaonic nuclei with up to six nucleons,  using sophisticated few-body techniques.  While such calculations produce binding energies that increase with the number of nucleons,  ranging from 25 MeV for $\bar{K}NN$ to about 80 MeV for $\bar{K}NNNNNN$,  the widths of these states are suggested to be of similar magnitudes but with large uncertainties, in particular because $\bar{K}$ multi-nucleon absorption processes are not taken into account.  For heavier antikaonic nuclei,  such $\bar{K}$ absorption reactions, when constrained by kaonic atom data and translated into a phenomenological potential with non-linear density dependence,  produce a large imaginary part of the potential that competes with the binding energy.  Thus the key issue in the physics of antikaonic nuclei that remains to be further clarified is the balance between nuclear $\bar{K}$ binding and absorption.  

%
%
%
%

%
\end{document}